\documentclass[12pt, preprint]{aastex}
\usepackage{rotating}
\def \lp{\>\> .}
\def \lc{\>\> ,}

\def \c2{cm$^{-2}$}

\def \kms{kms$^{-1}$}

\def \nh3{NH$_3$}
\def \n2h{N$_2$H$^+$}

\def \nh2{n_{H_2}}
\def \nh1{n_{HI}}

\def \tw{$^{12}$CO}

\def \h2{H$_2$}

\def \Ms{$M_{\odot}$}

\def \H2{H$_{2}$}

\def \be{\begin{equation}}
\def \ee{\end{equation}}
\def \bf{\begin{figure}}
\def \ef{\end{figure}}

\shorttitle{Cold HI Measurements in Molecular Cloud Cores}
\shortauthors{Kr\v{c}o et al.}

\begin{document}

\title{An Improved Technique for Measurement of Cold HI in Molecular Cloud Cores}

\author{Marko Kr\v{c}o}
\affil{Department of Astronomy, Cornell University, Ithaca, NY 14853}
\email{marko@astro.cornell.edu}

\author{Paul F. Goldsmith}
\affil{Jet Propulsion Laboratory, California Institute of Technology, Pasadena, CA 91109 and Department of Astronomy, Cornell University, Ithaca NY 14853}

\author{Robert L. Brown}
\affil{National Astronomy and Ionosphere Center, Cornell University, Ithaca, NY 14853 }

\and

\author{Di Li}
\affil{Jet Propulsion Laboratory, California Institude of Technology, Pasadena, CA 91109 }

\begin{abstract}
The presence of atomic gas mixed with molecular species in a ``molecular'' cloud may significantly affect its chemistry, 
the excitation of some species, and can serve as probe of the cloud's evolution.
Cold neutral atomic hydrogen (HI) in molecular clouds is revealed by its self absorption of background galactic HI 21-cm emission. 
The properties of this gas can be investigated quantitatively through observation of HI Narrow Self-Absorption (HINSA). 
In this paper, we present a new technique for  measuring atomic gas physical parameters from HINSA observations that  utilizes molecular tracers 
to guide the HINSA extraction.  
This technique offers a significant  improvement in the precision with which HI column densities can be determined over 
previous methods, and it opens several new avenues of study of relevance to the field of star formation.
\end{abstract}


\section{Introduction}

Star formation occurs in molecular clouds which are thought to have evolved from diffuse atomic hydrogen 
(HI) regions to form dense, cold, well--shielded regions composed primarily of molecular Hydrogen (\H2). 
Our quantitative understanding of cloud evolution and specifically the conversion process (from HI to \H2 gas) has been hindered by our inability to  measure confidently the HI abundance in evolving clouds. 
In this paper we present a new technique for measuring HI column densities in dark clouds offering a 
significant improvement over previous methods. 
In the interests of brevity this paper includes the results from only a few clouds on which this technique has been applied in order to demonstrate 
the technique. The results of a much larger survey of observational data and analysis are to follow in a subsequent publication.

The ability to determine accurately the HI component of molecular
clouds could have a variety of benefits. 
Measurement of  HI/\H2 ratios in clouds, used in
conjunction with astrochemical models, allows us to determine the
chemical ages of individual clouds or entire molecular complexes
(e.g. Taurus, Perseus, etc.). 
This will greatly expand our understanding by constraining star formation models thus yielding insights into the collapse process and the interplay of
magnetic fields, ambipolar diffusion, turblence, and various potential sources of cloud support. 
By studying age distributions in large scale regions we can learn about
the processes that may trigger the collapse of large complexes. 
In contrast to previous methods,  our technique allows us to determine the HI/\H2 ratios 
for individual velocity components within a cloud thus yielding unique information about cloud kinematics. 
Further, the technique allows for the absolute measurement of quantities such as HI column density, in contrast to many previous studies which were limited to comparative measurements.

HI is the dominant constituent of the diffuse ISM and its 21cm emission
line is prevalent everywhere throughout the sky, especially near the galactic
plane. 
Typical HI emission spectra are composed of numerous superimposed velocity components.  
The emission linewidths of the overall features are typically on the order of a few 10s of \kms. 
They include velocity variations owing to galactic rotation as well as very significant peculiar velocities resulting from localized phenomena.  
It is rare to find a molecular region for which one can confidently claim that the HI emission observed along the line of sight to the cloud is associated with the cloud, and is not due to background or foreground sources. 
Owing to such velocity crowding it is difficult or impossible to disentangle HI emission originating from a particular cloud from the background galactic HI emission.

The situation for HI absorption is different. 
HI within a galactic cloud of any type can absorb the continuum emission from distant (galactic or extragalactic) radio sources. 
Because the HI optical depth varies inversely with the cloud temperature, the absorption by galactic HI is stronger in cold, galactic HI clouds. 
The integrated optical depth of all the clouds along the line of sight through the galactic disk can exceed unity as demonstrated in \cite{Kolpak}, and \cite{Garwood}.

These cold, interstellar HI clouds may also be identified through their absorption of warm background HI emission originating within the galaxy. 
Because the emission being absorbed by cold HI clouds in this case is galactic HI emission, the resultant spectral absorption features are refered to as HI self-absorption (HISA). There have been many surveys with HISA detections over the years including \cite{garzoli}, \cite{Heiles}, \cite{Knapp}, \cite{Heiles2}, \cite{Wilson}, \cite{McCutcheon}, \cite{Myers}, \cite{Bowers}, \cite{Batrla}, \cite{Shuter}, \cite{vanderWerf}, \cite{Feldt}, \cite{Montgomery}, \cite{Gibson}, and \cite{Kavars03}, among others. 
It is important to emphasize that the term HISA refers to an observable spectral absorption feature rather than being a description of a specific physical process.

Molecular spectral emission lines provide an independent view of that subset of interstellar clouds that are cold and composed primarily of molecular species. 
These ``molecular clouds'' are expected to maintain a residual abundance of atomic hydrogen if for no other reason than the cosmic ray disassociation of \H2. 
This applies even when the chemical evolution of the cloud has reached equilibrium \citep{Solomon1971, HINSA2}. 
With the residual HI co--existing throughout the cloud with molecular species, the observed spatial and velocity (kinematic) structure of the molecular cloud will be similar whether observed in molecular emission or HI self-absorption lines. Thus, we expect to observe HI self-absorption features along the lines of sight of molecular clouds which share the spatial distribution and kinematics (non--thermal line width) of molecular emission lines. 
Such localized association between molecular emission and HI self-absorption is observed for many nearby clouds as reported by \cite{HINSA1} and \cite{HINSA2}. 
The specific case in which the HI absorption features observed in the direction of a molecular cloud share the spatial and kinematic structure seen in the molecular lines is called HI Narrow Self-Absorption (HINSA) as defined in \cite{HINSA1}. 
The term {\it Narrow} arises from the typically small nonthermal linewidths (on the order of 0.1 \kms) of HINSA features, very similar to the similarly small nonthermal linewidths of molecular tracers along the same lines of sight. 
HINSA can be considered to be a subset of HISA, but it is a subset derived from an understanding of a specific, observable physical phenomenon in molecular clouds.

While both are simply acronyms for spectral absorption features, HISA can be caused by a variety of different conditions and processes, but HI Narrow Self-Absorption (HINSA) is that subset of HISA in which the atomic HI absorption correlates well with molecular emission of certain tracers (most notably $^{13}$CO) in sky position, central velocity, and nonthermal line width. 
Based on our current understanding of cold molecular clouds, the most satisfactory picture is that HINSA features are a result of HI gas located within these cold, dense, well--shielded regions. 
Some early examples of HINSA studies that predate the use of this term are those of \cite{Wilson}, \cite{vanderWerf}, and \cite{Jackson}. 
The technique which we describe in this paper pertains only to the extraction of HI data from HINSA features.

The general picture which emerges, as found in \cite{HINSA1} and \cite{HINSA2}, is that the HI gas located within cold, quiescent cores of dark clouds produces HINSA absorption features.  
The well--defined center velocities and narrow line widths allow us to separate the HI gas associated with individual clouds from the galactic background. 
However, the complexity of the background emission spectra that are frequently encountered 
makes extracting accurate data (especially in terms of obtaining the cold HI column density) from the absorption features difficult.  
Several methods (discussed in \S \ref{technique}) have been used previously, but all are recognized to  introduce significant uncertainties in the results. 
We here present a new technique that aims to improve the situation by using the properties of molecular emission to characterize the region producing the HINSA features, and then employes the HINSA spectral features to derive HI column densities.

In \S \ref{real} we present selected data and show the results of applying the new technique to them, and in \S \ref{previous} we contrast this with previous methods for analyzing HINSA data. In \S \ref{technique} and \S \ref{beamsizes} we describe the technique and the combination of molecular data with the HINSA spectra. In \S \ref{examples} and \S \ref{ambiguity} we verify the validity of our technique using simulated data and also examine its limitations.

\section{Results Using Observational Data}
\label{real}

In this section we demonstrate the utility of our procedure by showing some results of its implementation based on observations of two molecular clouds. 
While a full discussion of the implications of these results is reserved for another publication, we hope that the results obtained encourage careful consideration of the technique described here.

\begin{figure}
\plotone{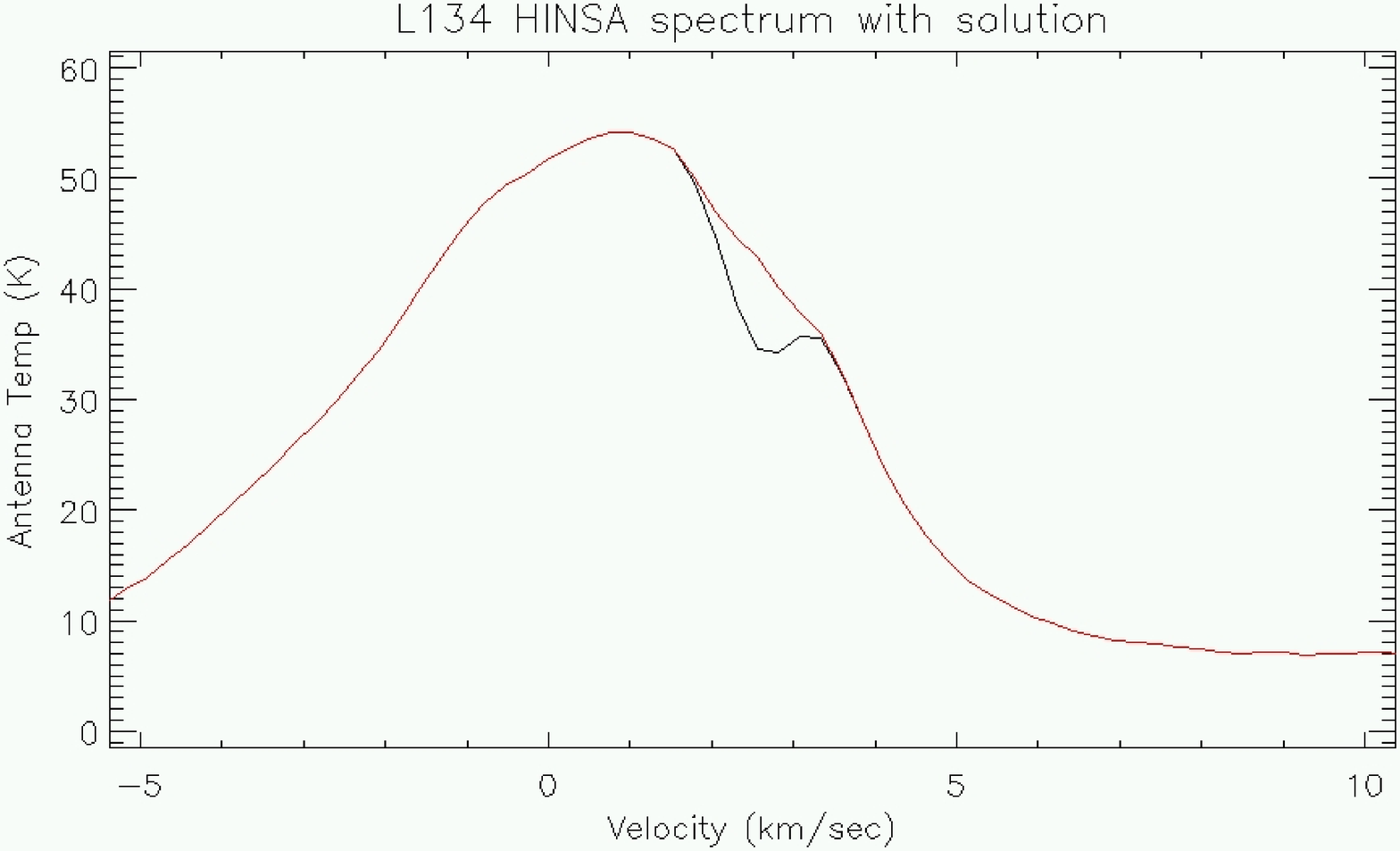}
\plotone{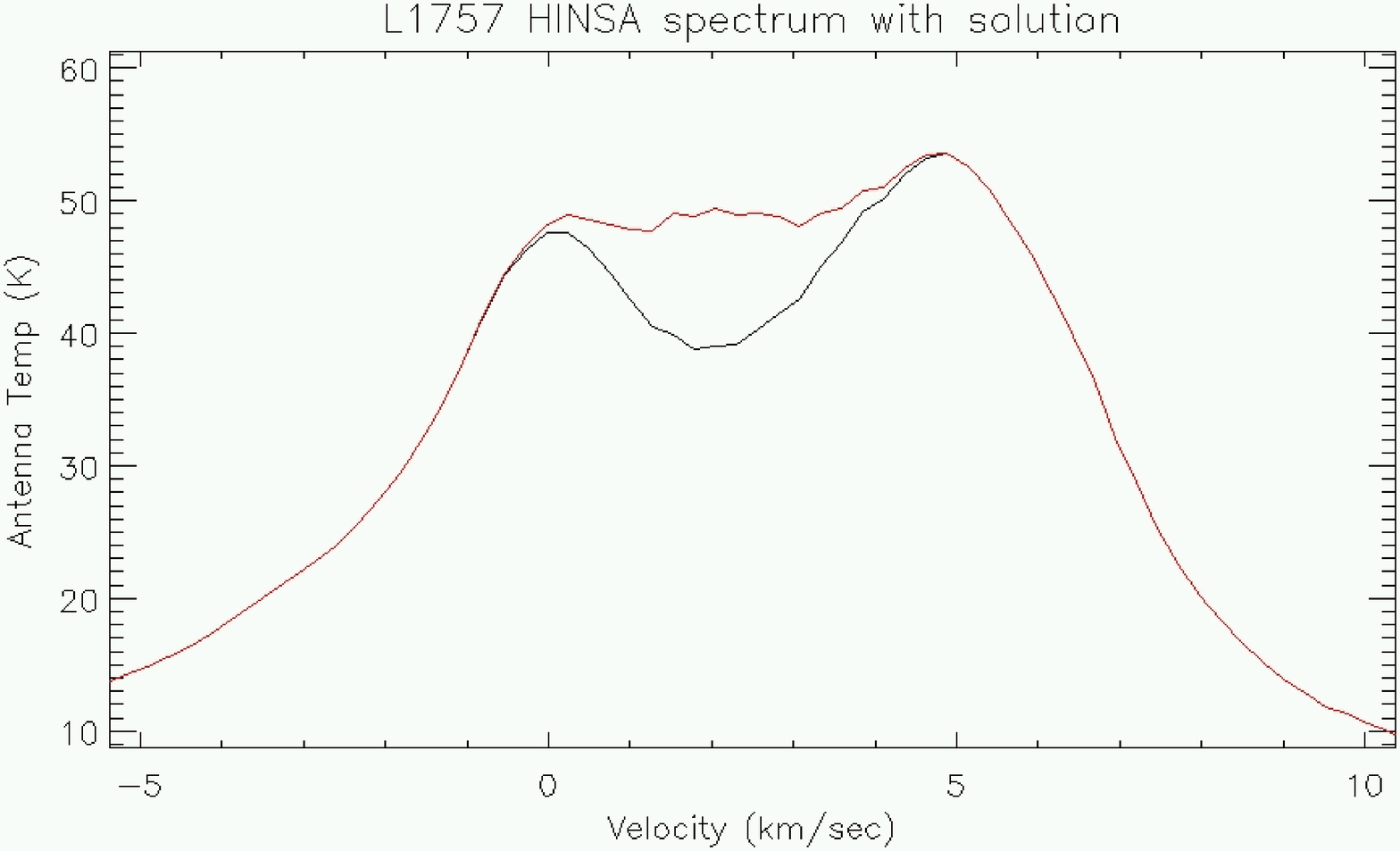}
\caption{
Sample HI spectra in the directions of the mlecular clouds L134 (upper) and L1757 (lower). We show the original observed spectra (black) and the recovered background spectra after HINSA removal (red), using the technique described in this paper.  
\label{HINSAspec}
}
\end{figure}

An observational survey, whose complete results and analysis are presented elsewhere, was performed using the Green Bank Telescope (GBT) and the Five College Radio Observatory (FRCAO) to obtain HI, OH, $^{13}$CO, $^{12}$CO, and C$^{18}$O maps of more than 30 dark clouds which exhibit
HINSA features. 
Figure \ref{HINSAspec} shows sample HI spectra from two well known clouds, Lynds 134 (L134) and L1757. 
HINSA absorption features are prominent in both sources. 
The red lines represent the recovered background spectra after HINSA has been removed. 
This representation is reassuring in that the smooth, natural looking background
spectra indicate that the HINSA removal process has not produced marked distortions in the background.  
While L134 represents a simple case of a single emission component, in L1757 our technique is
able to discern two emission components closely spaced in velocity. 
Previous methods based solely on the HI line profile would have significantly overestimated the HI optical depth by assuming there was only a single emission component. 
By utilizing the associated molecular data as described below, the possibility of a single emission component is largely eliminated.

The technique we use here assumes that the cold HI and molecular constituents are well--correlated spatially, so that we can assume that the line center velocity and nonthermal line width for the cold atomic gas and the molecular gas in a particular direction are the same. 
This is justified by previous studies of HINSA and CO isotopologues, including \cite{HINSA1} and \cite{HINSA2}.
Our technique depends on use of these spectra to calculate the HI optical depths and column densities. 
In conjunction with observations of molecular tracers this allows us to obtain HI column densities in dark clouds with improved accuracy. 
Further, we are able to obtain unprecedented detail by examining individual velocity 
components within each cloud, even when their velocity separations are small enough to make individual components difficult to distinguish in HI spectra. 
Additional details are given in \S \ref{technique}

\begin{figure}
\epsscale{0.65}
\plotone{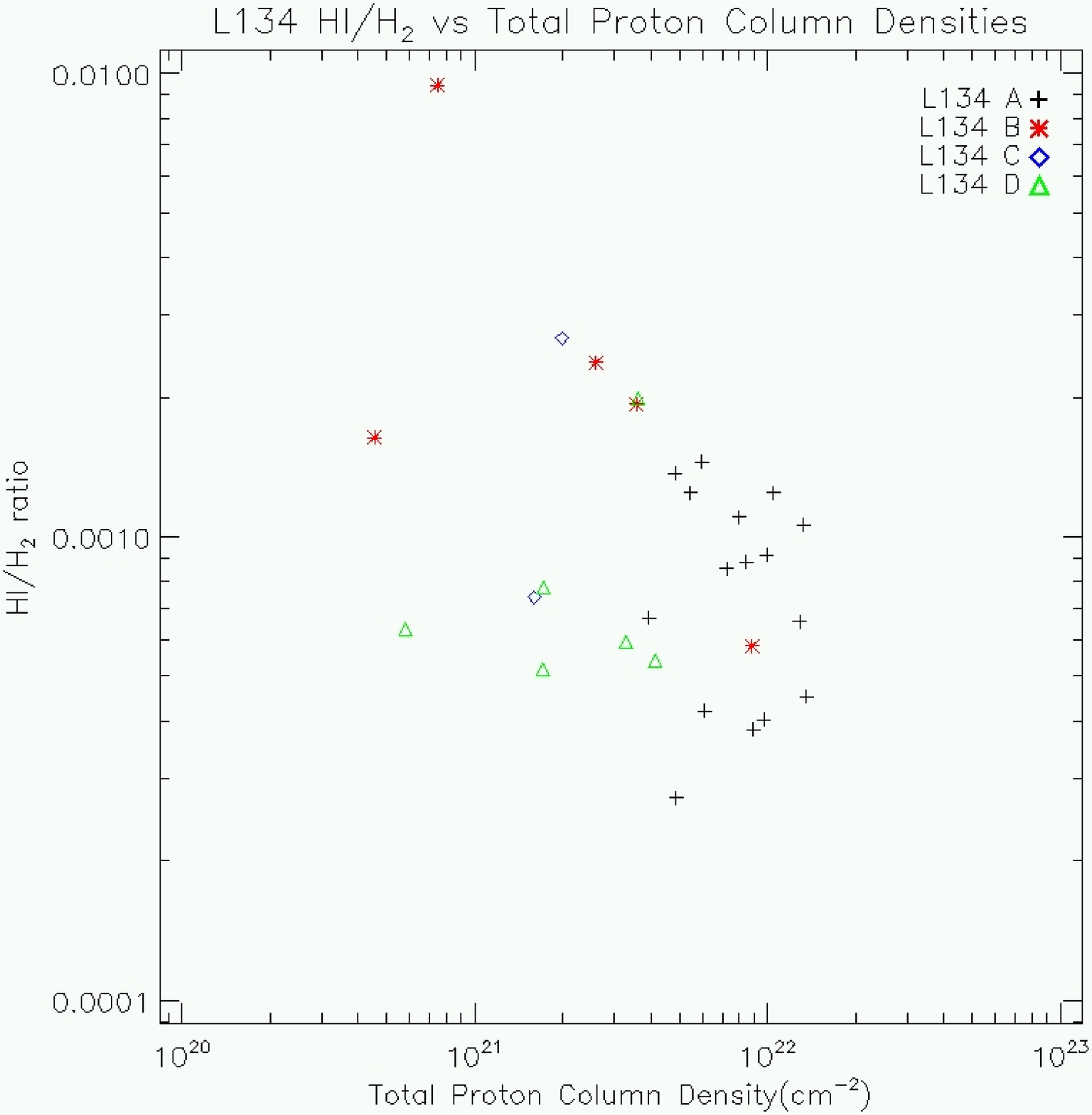}
\plotone{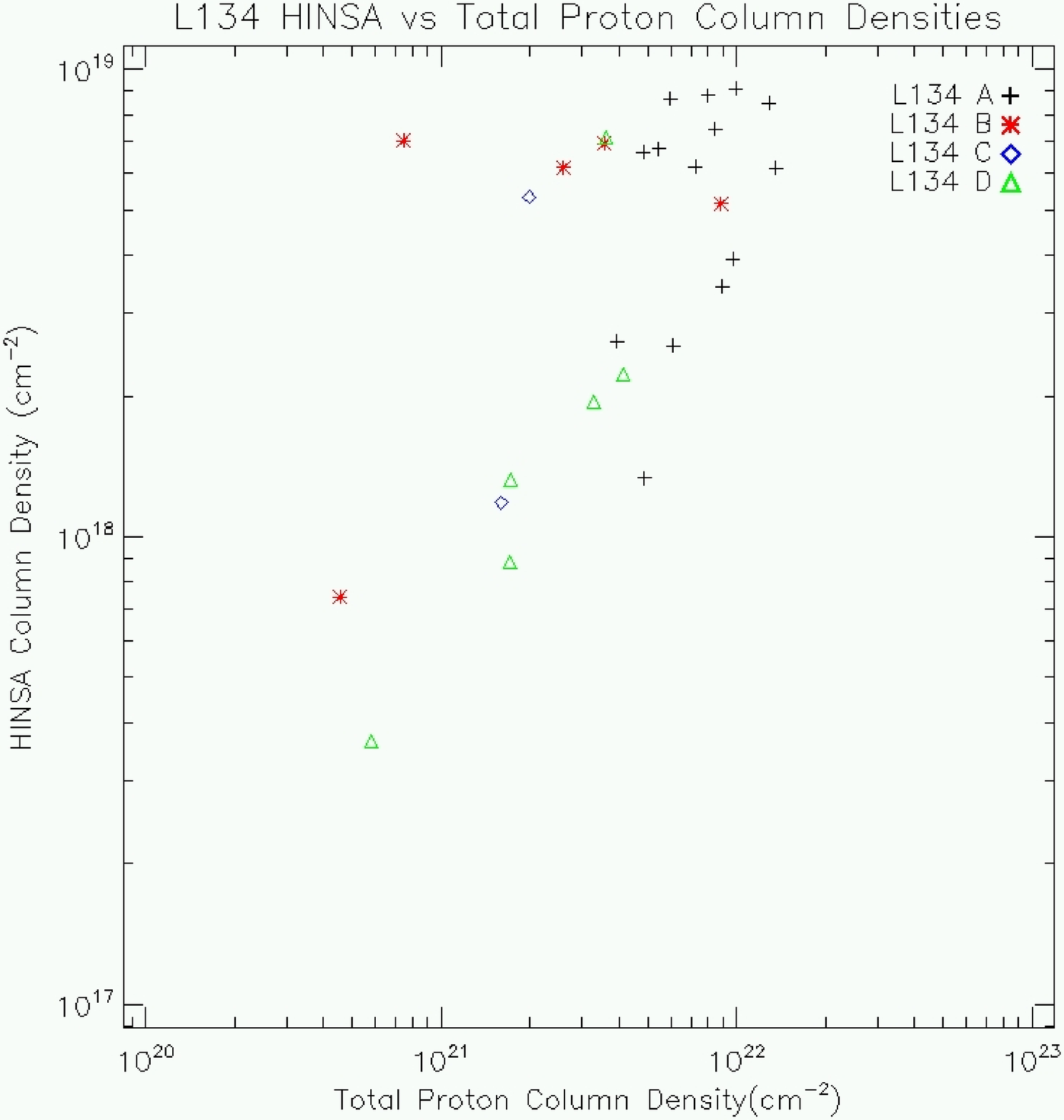}
\caption{
Atomic to molecular abundance ratios in the molecular cloud L134. Each point represents a single velocity component at different position within the cloud. The 2.55--3.5, 1.65--2.55, 1.55--1.65, 0.25--1.55, and -1.00--0.25 \kms\ velocity components are represented by black, red, blue, green, and cyan dots, respectively. These points are Nyquist sampled relative to the GBT HI beam and are thus independent.  There are no systematic differences in the HI/H$_2$ ratio among the different velocity components.
\label{HINSAplots}
}
\end{figure}

\begin{figure}
\epsscale{0.65}
\plotone{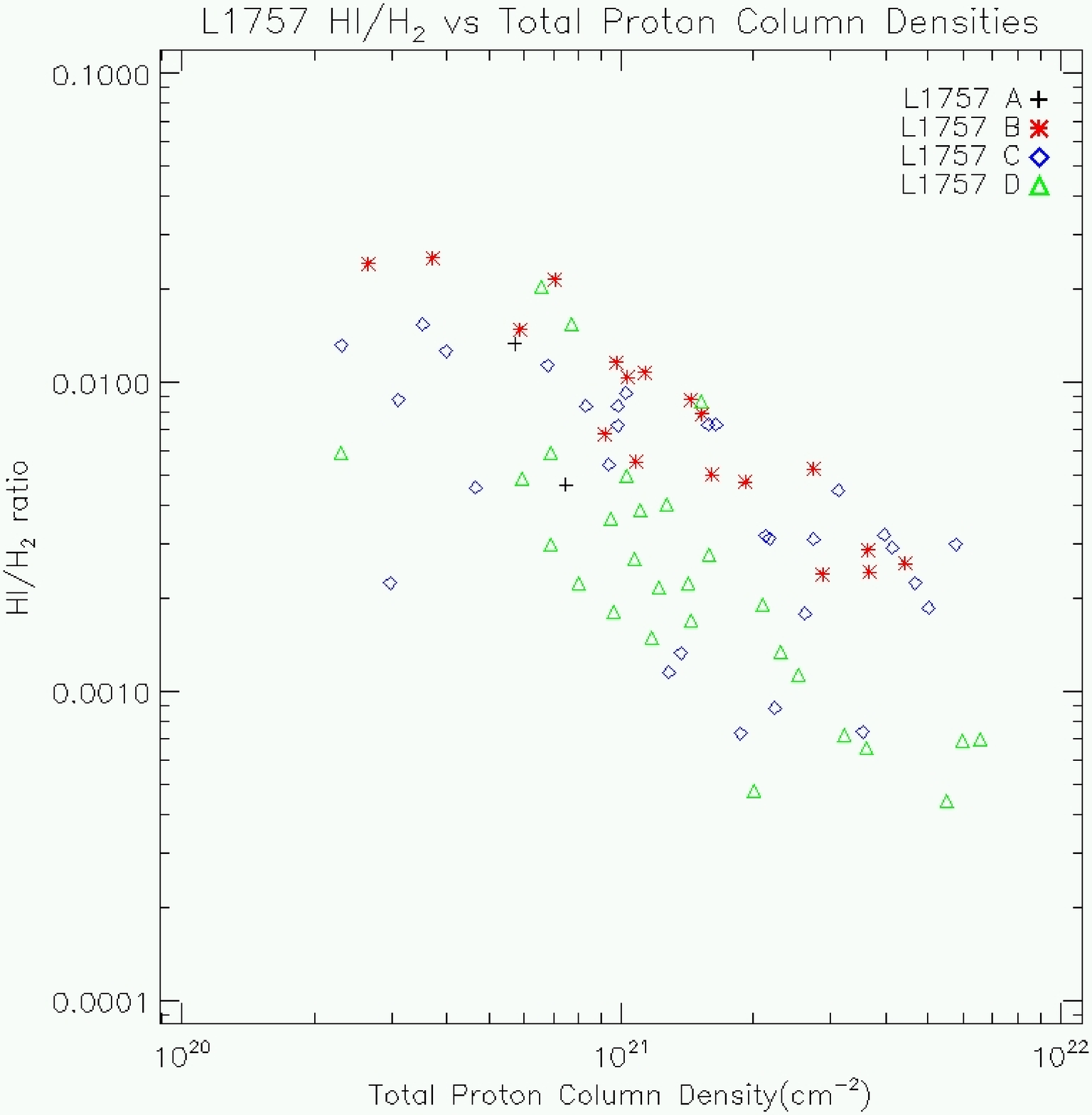}
\plotone{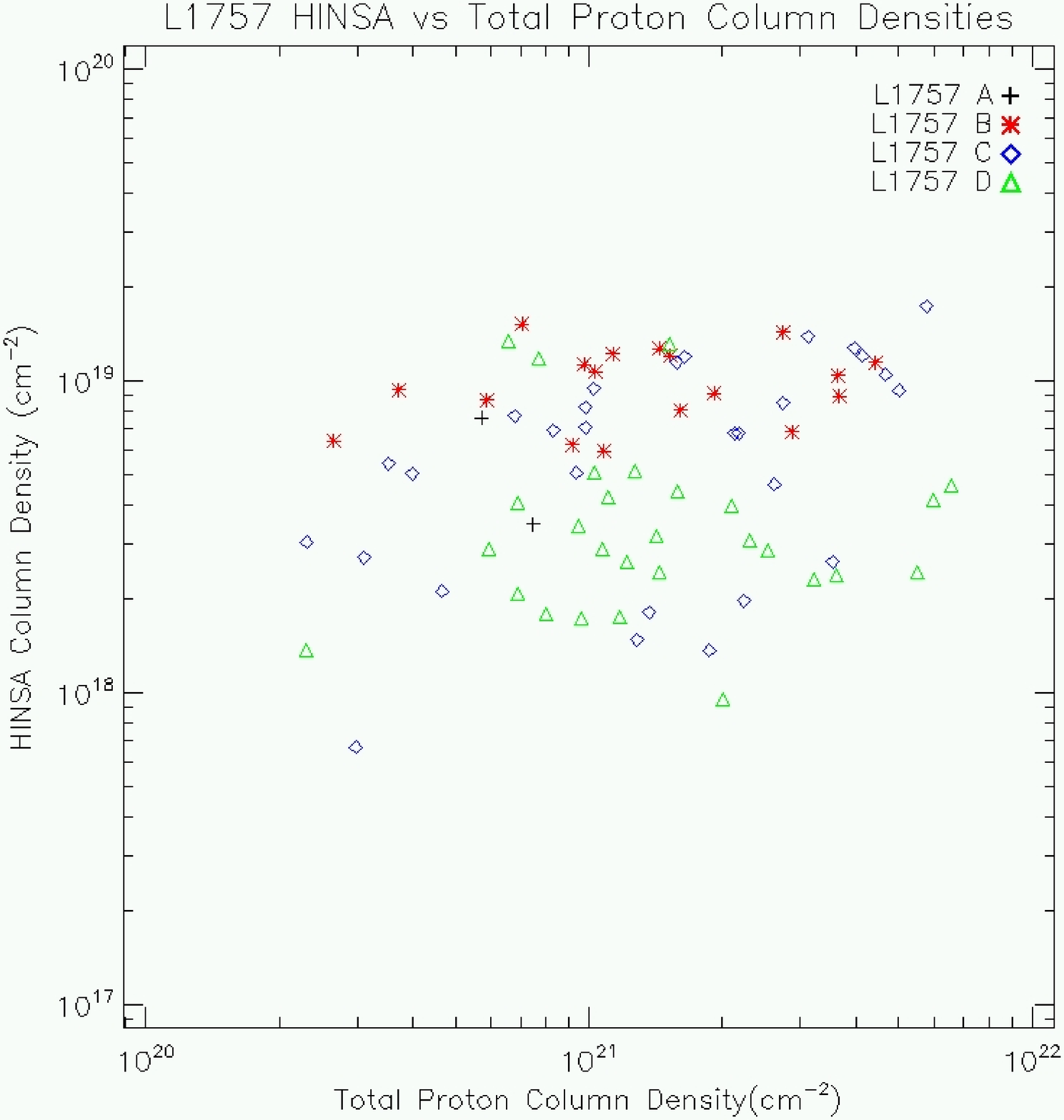}
\caption{
Atomic and molecular hydrogen column densities and ratios in L1757, presented  as in Figure \ref{HINSAplots}. The 4.3--6.7, 2.0--4.0, 1.0--2.5, -0.5--1.3 \kms\ components are represented by black, red, blue, and green points, respectively. Unlike L134, the various velocity components of L1757 show fairly well--defined but systematically different different relationships between HI and H$_2$. 
\label{HINSAplots2}
}
\end{figure}

Figures \ref{HINSAplots}, \ref{HINSAplots2}, and Tables \ref{HINSAtable} and \ref{HINSAtable2} show measured HI, H$_2$, and total proton column densities, and the HI/H$_{2}$ ratios for individual velocity components throughout numerous positions in clouds L134 and L1757. 
No such analysis has previously been carried out. 
The choice of these two sources indicates the importance and possible rewards of such studies in that the two clouds exhibit very different properties. 
While in L134 all the velocity components have similar HI/H$_{2}$ abundances in the relatively narrow range between 10$^{-4}$ and 10$^{-2}$, in L1757 the situation is quite
different, with ratios varying systematically as a function of velocity and extending over a much greater range. 
The most likely explanation for this phenomenon is that L134 is a mature cloud in which the chemical evolution of all of the velocity components has proceded to the point where HI abundances have approached their equilibrium values.  
In contrast, L1757 represents a much younger cloud in which each component is evolving at a different rate governed by its total density. 
This information yields valuable insight into the dynamics and chemistry of dark clouds, but has never before been available. 
A complete discussion of the results for a large sample of clouds will be presented in a forthcoming paper as a proper analysis requires more space than is available here.

\begin{sidewaystable}[h]
\begin{tabular}{|c|c|c|c|c|c|c|}
\hline
 & $V_{LSR}$(\kms) & HI Mass(\Ms) & H$_2$ Mass(\Ms) & Mean HI/H$_2$ & Peak H$_2$ Dens.(cm$^{-3}$) & Mean H$_2$ Dens.(cm$^{-3}$)\\
\hline
L134 A & $2.35-3.40$ &$4.0 \cdot 10^{-2}$ & $71 $ & $1.1 \cdot 10^{-3}$ & $2490$ & $910$\\
L134 B & $1.80-2.75$ &$9.5 \cdot 10^{-3}$ & $41 $ & $4.7 \cdot 10^{-4}$ & $2390$ & $660$\\
L134 C & $0.50-2.75$ &$4.7 \cdot 10^{-3}$ & $8.4$ & $1.1 \cdot 10^{-3}$ & $1530$ & $310$\\
L134 D & $-0.7-1.10$ &$5.1 \cdot 10^{-3}$ & $12$  & $8.5 \cdot 10^{-4}$ & $1230$ & $480$\\

\hline
L1757 A & $4.70-6.00$ &$1.6 \cdot 10^{-2}$ & $31$ & $1.0 \cdot 10^{-3}$ & $840$ & $310$\\
L1757 B & $2.50-3.35$ &$2.7 \cdot 10^{-1}$ & $45$ & $1.1 \cdot 10^{-2}$ & $540$ & $270$\\
L1757 C & $1.50-2.20$ &$3.0 \cdot 10^{-1}$ & $106$& $5.7 \cdot 10^{-3}$ & $1560$& $305$\\
L1757 D & $0.80-1.50$ &$1.7 \cdot 10^{-1}$ & $95$ & $3.5 \cdot 10^{-3}$ & $880$ & $290$\\
\hline
\end{tabular}
\caption{Derived Component Properties for L134 and L1757. The component radii are estimated by the extent of the $^{13}$CO emission and are used to estimate cloud depths and thus volume densities. In L1757, the fact that some components have roughly the same H$_{2}$ density but very different HI abundances suggests that the different components may have formed at different times. 
\label{HINSAtable}
}
\end{sidewaystable}

\begin{sidewaystable}[h]
\begin{tabular}{|c|c|c|c|c|}
\hline
 & Peak HINSA optical depth & Peak $^{13}CO$ optical depth & Peak $^{12}CO$ T$_{ex}$ (K) & Mean continuum at 21cm (K)\\
\hline
L134 A & 0.53 & 0.84 & 19 & 3.1\\
L134 B & 0.20 & 0.67 & 17 & ~\\
L134 C & 0.20 & 0.24 & 17 & ~\\
L134 D & 0.19 & 0.22 & 14 & ~\\

\hline
L1757 A & 0.20 & 0.31 & 16 & 3.2\\
L1757 B & 0.35 & 0.29 & 17 & ~\\
L1757 C & 0.36 & 0.74 & 19 & ~\\
L1757 D & 0.30 & 0.35 & 15 & ~\\
\hline
\end{tabular}
\caption{Derived Component Properties for L134 and L1757 continued. The $^{12}$CO temperature, derived directly from the $^{12}$CO spectra, is used as an estimate of the HI temperature represented by $T_H$ in Equation \ref{maineqn}. 
\label{HINSAtable2}
}
\end{sidewaystable}

\section{Limitations of Previous HI Self-Absorption Analysis Techniques}
\label{previous}
Owing to the complexity of HI background emission along most lines of sight, it is clearly 
quite arbitrary to fit a straight line across the absorption feature. 
It is equally problematic to use OFF source observations since the HI background spectra 
change significantly over the angular size of the foreground cloud as sampled by the telescope beams that have been employed. 
Several methods for extracting HI column densities from HISA absorption have been tried in the past. 
These can be classified into two groups, each with its own set of limitations.
The most commonly--used methods rely on fitting simple mathematical functions such as 
gaussians or polynomials over the observed spectra while masking the absorption features. 
As discussed in \cite{HINSA1} these techniques result in significant uncertainty because the analysis is under-constrained.  
Arbitrary choices for the degree of the fitted polynomial and the masking ranges must be made, and these greatly affect the final results.

Recently, \cite{Kavars05} implemented an automated algorithm for measuring HISA which relies on spotting the characteristic HISA features (the dip) in HI spectra. 
However, this technique (as pointed out by its authors) suffers from several limitations. 
The \cite{Kavars05} technique, as well as function fitting methods, are unable to determine the gas temperature and the optical depth without assuming a value for one of these important quantities. These methods are thus limited to comparative measurement of one quantity, such as HI column density, over large regions recognizing that absolute values cannot be obtained.

Owing to the complexity of the galactic HI background, and lacking additional information,  it is often impossible to tell which dips in the spectra are genuine absorption features and which are produced simply by the superposition of two emission features. 
Both phenomena may be spatially extended with continuous velocity distributions.  
Molecular observations can be used to distinguish between these two scenarios.
In molecular clouds there are often multiple components along the same line of sight with closely spaced central velocities. 
The lower mass of atomic hydrogen compared to molecular tracers (and consequently greater thermal line width)  means that it is often possible for multiple components which are distinguishable in molecular spectra to merge into a single HI absorption feature. 
Using the molecular gas as a template for the HINSA analysis has the potential to greatly enhance the accuracy of the HI/H$_{2}$ comparison since the ratio may be different in each velocity component.

Previous analysis techniques have been applied to both HISA and HINSA sources
without distinguishing between the two. 
Our technique is limited to HINSA sources for which (by definition) there is molecular data available. 
Limiting our scope in this manner allows us to utilize observational data from various molecular tracers to assist in extracting meaningful results from HINSA features. 
As will be shown in what follows, to the extent that our assumption that HI is mixed with the molecular gas is valid, we are able to measure both the temperature and optical depth of the HI gas without any ambiguities regarding fitting functions. In this way we may treat each velocity component in a cloud separately.

\section{The Technique}
\label{technique}
\subsection{Analytic Representation}
\label{techniqueanalytical}
An idealized HINSA spectrum can be represented by the radiation transfer equation as

\begin{equation}
\label{maineqn}
T_{A}(v)=(T_{b}(v) + T_{c})e^{-\tau(v)} + T_{H}\left( 1 - e^{-\tau(v)}\right) \lc
\end{equation}
where $v$ corresponds to velocity, $T_{A}(v)$ is the observed spectrum, $T_{b}(v)$ is the
background emission due HI emission clouds, $T_c$ represents the continuum emission along the line of sight at 21 cm including the 2.7K microwave background(\S \ref{HIcont}), $T_{H}$ is the gas temperature of the foreground absorbing cloud producing HINSA, and $\tau(v)$ is the optical depth of the absorbing HI gas at 21cm which we can describe as
\begin{math}
\tau(v)=\tau_{0}e^{\frac{- (v - v_{H})^2}{2\sigma_{H}^2}} \lc
\end{math}
where $\tau_{0}$ is the peak optical depth, $v_{H}$ is the central
velocity of the HINSA component, and $\sigma_{H}$ is the HINSA
linewidth. 
In this particular representation we have assumed that there is no intervening HI emission originating between us and the absorbing HINSA cloud; this topic is further discussed in \S \ref{HIfore}. When multiple absorption components overlap in velocity within the same line of sight,
the transfer equation must be applied consecutively for each component from farthest to nearest.

HINSA absorption features are typically gaussian-like dips with much smaller amplitudes and linewidths than the characteristics of the emission spectra ($T_b(v)$) on which they reside.  These features can be so small in comparison to the background emission that they can be difficult to distinguish by eye. However, by
utilizing a fundamental property of gaussians we can obtain a much better representation of the data in which the HINSA becomes more apparent.

\begin{figure}
\plotone{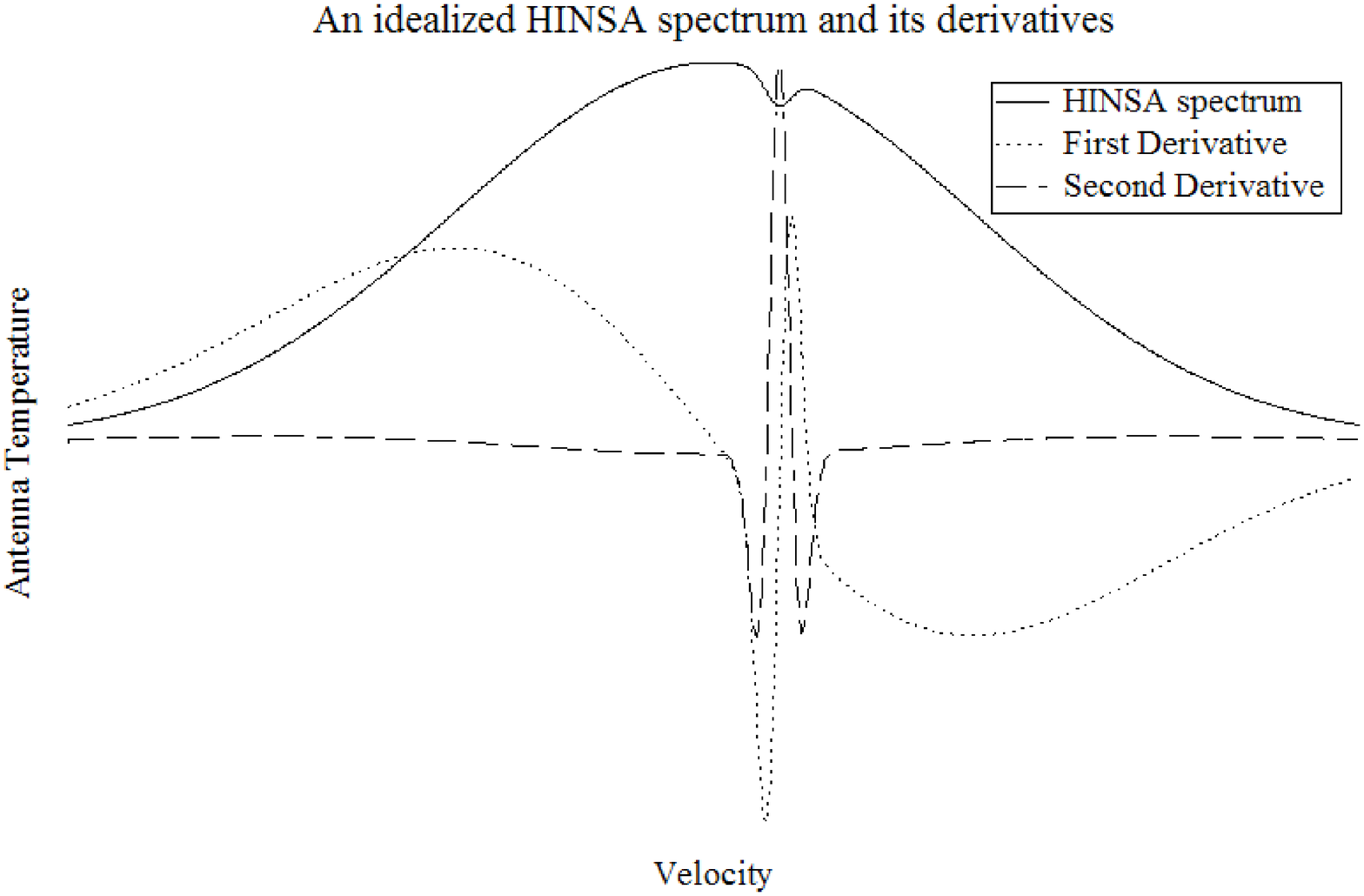}
\plotone{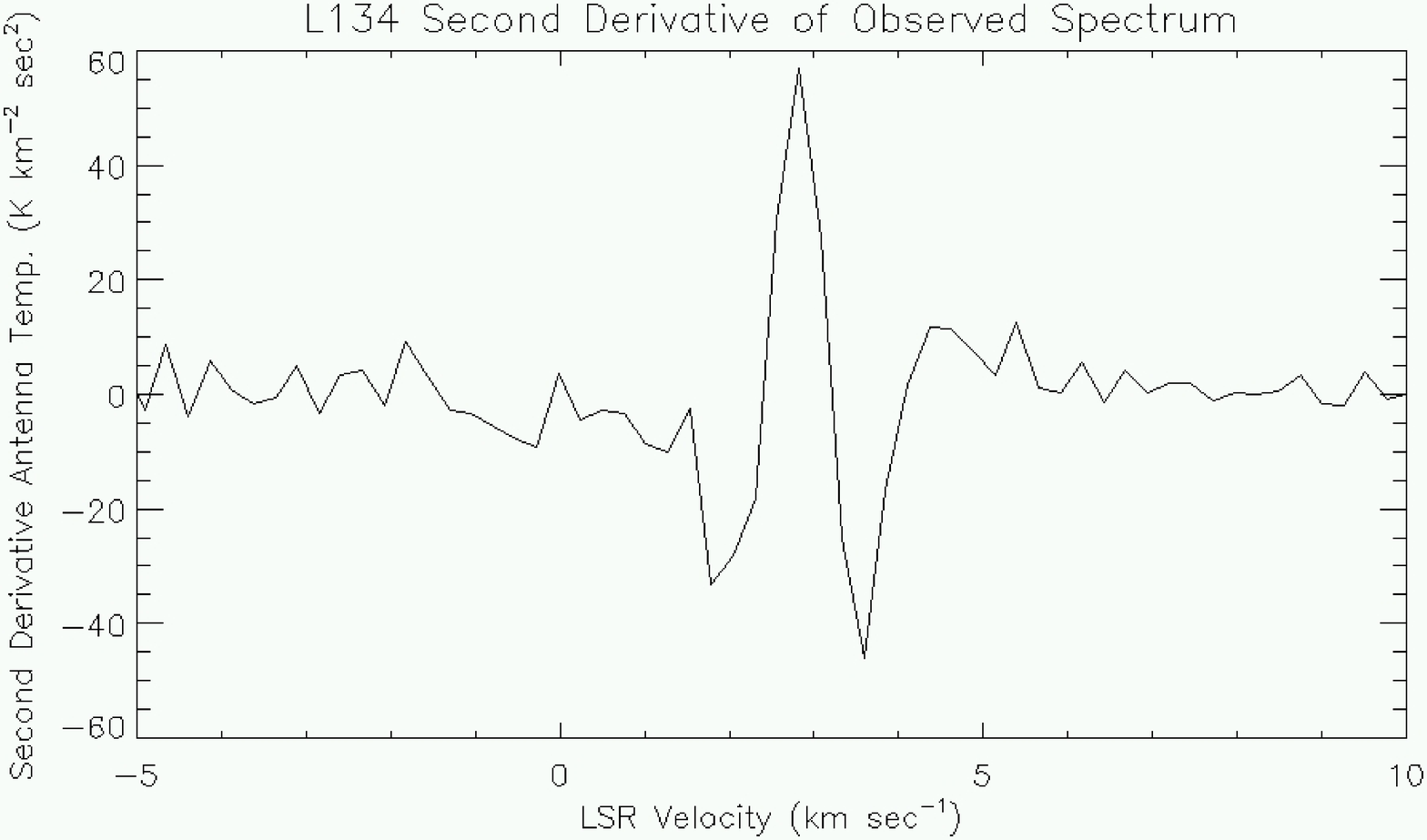}
\caption{(upper) An idealized HINSA spectrum with typical observed properties along with its first and second
  velocity derivatives. The scales have been normalized for
  demonstration purposes. (lower) Second derivative representation of the observed spectrum toward L134 shown is Figure \ref{HINSAspec}. It is apparent that a small HINSA dip
  becomes the dominant feature in the second derivative
  representation\label{idealized}}
\end{figure}

The derivatives in velocity or frequency of a gaussian function can be written as:

\begin{equation}
\begin{array}{ccc}
f(v)=f_{0} e^{(-v^2/2\sigma^2)} & \frac{df(v)}{dv}=-\frac{v}{\sigma^2}f(v) &\frac{d^{2}f(v)}{dv^2}=- \frac{1}{\sigma^2}f(v) + \frac{v^2}{\sigma^4}f(v) \lc
\end{array} 
\end{equation}

where it is apparent that higher order derivatives show a stronger
dependence on linewidth. For conceptual purposes we can temporarily
approximate the HINSA absorption and background emission features as gaussians. 
Because the linewidths for HINSA features (typically
~1 \kms) are much narrower than the linewidths for the
background emission components, in the first and second
derivative representations, even a low-amplitude absorption feature will
become readily apparent in comparison to the bright background emission. This is
illustrated in Figure \ref{idealized}. In the second derivative
representation, HINSA becomes the dominant feature in the spectrum thus making it easier to isolate and quantify.

One could in principle use higher derivative representations to
search for the best fit as the HINSA becomes progressively more
prominent. however with real data the noise also rises with higher
derivatives. The second derivative representation has been found to be the optimum choice for our GBT and Arecibo observations.
It would be possible to take the second derivative representation of any HINSA spectrum and 
attempt to fit a simple function to it to extract the absorption. 
However, HINSA spectra are rarely so simple as to lend themselves to this kind of fitting and we 
would have very limited information on the temperature of the HI gas. 
Furthermore, the large thermal linewidths of HI spectra tend to blend all the velocity components 
along a single line of sight. 
Unless we are convinced that all the velocity components in a cloud can be characterized by the same HI/H$_{2}$ 
ratio, we must find a way to extract the HINSA using additional constraints based on physical parameters. 

The primary benefit of limiting our technique to HINSA features is
that we can use data from molecular tracers in order to constrain the analysis. 
If we can determine the values of all the other parameters (linewidth,
center velocity, and temperature) for each velocity component then we
can ideally limit ourselves to searching for that value of $\tau_{0}$ which produces the $T_{b}$ emission spectrum whose second
derivative best removes the HINSA feature. Minimizing the integrated squared sum of the second derivative
background spectrum is sufficient to extract the HINSA feature (see \S \ref{fitting}).

\subsection{Continuum Emission at HI wavelengths}
\label{HIcont}
    There are various sources of continuum emission (T$_c$) in equation \ref{maineqn} throughout the galaxy that contribute to the sky brightness at 21cm. The emission may originate from a variety of sources such as synchroton emission within the galaxy and continuum emission from ionized gas such as that in the Galactic Halo. Continuum emission can also originate from individual discrete galactic and extra-galactic sources,  such as pulsars, HII regions, AGNs, accretion disks, and quasars. These sources will typically produce emission whose intensity is constant across the observed HI band. Observing techniques such as frequency switching as well as software baseline removal algorithms remove the continuum emission and the 2.7K background along a particular line of sight from the reduced spectrum. The continuum emission component should ideally be included when calculating HINSA since the temperature of the cold gas clouds (~10K) is comparable to the antenna temperature produced by the continuum emission. Several galactic surveys have been performed which have measured the value of the 21cm continuum including \cite{Reichetal}, \cite{Uyaniker}, and \cite{ReichReich}, among others. These data can be utilized to determine the expected continuum emission along any HINSA source. Typical values are on the order of a few K. According to equation \ref{maineqn}, the value of T$_c$, while significant, becomes critical only when the HINSA optical depths are high and gas the temperature is low.

\subsection{Foreground HI Emission}
\label{HIfore}
    Usually there will be HI emission between the observer and the HINSA cloud due to intervening HI gas. To assess the importance of the effect this invervening gas might have on our HINSA measurements we must expand equation \ref{maineqn} to include all three gas components, that of the HI emission behind the HINSA cloud, absorption due to the HINSA cloud itself, and emission between us and the HINSA cloud.

\begin{equation}
\label{TI}
T_I (v) = T_c e^{-\tau_{b}(v)} + T_{ex~b}(1 - e^{-\tau_{b}(v)})
\end{equation}
represents the emission component due to the warm HI gas behind the HINSA cloud whose excitation temperature is $T_{ex~b}$ and optical depth is $\tau_b(v)$. At this stage, only the background continuum emission $T_c$ is attenuated.

The HINSA cloud then attenuates the background emission, while also adding its own contribution as in
\begin{equation}
\label{TII}
T_{II} (v)= T_I(v) e^{-\tau_H (v)} + T_{ex~H}(1-e^{-\tau_{H}(v)})~,
\end{equation}
where $\tau_H$ and $Tex_H$ represent the optical depth and excitation temperature of the cold HI gas within a molecular cloud. Similarly, any foreground gas will attenuate the spectrum further as well as adding its own emission to produce the final observed spectrum:
\begin{equation}
\label{TIII}
T_{III} (v) = T_A(v)= T_{II}(v) e^{-\tau_f(v)} + T_{ex~f}(1-e^{-\tau_{f}(v)})~,
\end{equation}
where $\tau_f$ and $T_{ex~f}$ represent the optical depth and excitation temperature of the foreground HI gas between us and HINSA cloud.

We now have two new terms, $\tau_f$ and $T_{ex~f}$, describing emission due to HI gas between us and the HINSA cloud. There is no way to operationally disentangle which portion of the observed emission spectrum originates from in front or behind the molecular cloud, but clearly any foreground emission we observe will attenuate the HINSA features and we must somehow correct for this. To get a sense of the effect and scale of the attenuation, we can assume that both the background and foreground emission gas have the same excitation temperature $T_{ex~tot} = T_{ex~b} = T_{ex~f}$. We can further assume the HI gas is distributed uniformly throughout the galaxy along the line of sight so that a linear relationship can be established between $\tau_b$ and $\tau_f$ such that $\tau_b = p \tau_{tot}$ and $\tau_f =(1-p)\tau_{tot}$ where $\tau_{tot}$ is the total optical depth of all the emitting HI gas behind and in front of the HINSA cloud and $p$ represents the fractional distribution between the two which we can estimate using
\begin{equation}
\label{pequation}
p(\emph{b}) = \left(1 - \frac{D_{cloud} sin(\emph{b})}{R_{disk}}\right)~.
\end{equation}
Here, $D_{cloud}$ is the distance to the HINSA cloud, $\emph{b}$ is the galactic longitude along the given line of sight, and $R_{disk}$ represents the distance from Earth to the edge of the galactic disk as defined by its HI content. With a typical scale height of 100pc near the Sun's position within the Milky way we take this value to be ~200pc \citep{stahler}. For a more complete discussion see \cite{vanderWerf} or \cite{HINSA1}. In this case we are using geometry of the local galactic disk, and the HINSA cloud's position within the disk to estimate how much of the HI emission visible along a particular line of sight is in fact located in front of a HINSA source attenuating the HINSA feature.

This emission, if located at the same velocity as the HINSA feature, will have the effect of washing out the absorption feature and thus the apparent HINSA optical depth will be lower than the intrinsic value. In fact, it can be shown from the above formulation that for optically thin HINSA, $p(\emph{l,b}) \propto \tau^{-1}$ as showin in \cite{vanderWerf}, where $\tau$ now represents the observed apparent optical depth of any HINSA feature. Thus to obtain the intrinsic HINSA optical depth of a cloud we must divide the apparent optical depth by $p(\emph{l,b})$. Due to the uncertainties in the HI distribution it is impractical to directly model the effects of foreground clouds when trying to extract HINSA profiles. Instead HINSA optical depths are computed on the assumption that there is no intervening emission and are subsequently adjusted using the factor $p(\emph{l,b})$ which itself depends only on the cloud's position and is calculated independently of the observed emission spectrum. 

The observed spectra thus give us an apparent optical depth which then must be adjusted by p to yield an estimate of the intrinsic optical depth.
For most nearby clouds this correction factor is in fact small in comparison to other sources or error. It must be true that $p(\emph{l,b})$ is not much less than unity if a significant HINSA feature is to be observed otherwise the foreground emission would obscure the HINSA absorption. For the two sample clouds in this paper, L134 and L1757 the correction factor has values of 0.5 and 0.7 accordingly.

\subsection{Obtaining Constraints Using Molecular Parameters}
\label{constraints}
It is possible to obtain data from molecular tracers which can provide
us with reliable estimates of the HINSA feature properties for each
velocity component by utilizing certain assumptions justified by previous work.
\cite{HINSA1} showed a very strong correlation between the line center
velocity of each HINSA component and those of its molecular
counterparts using the OH, $^{13}$CO, and C$^{18}$O molecules as tracers. 
Thus, once each molecular component has been identified and separated
using traditional fitting we can assume that each velocity component
visible in molecular emission has an accompanying HINSA component whose optical depth may yet turn out to be 0. We can use the central velocity from any of the above molecules as an estimate of the number of HINSA components present and their center velocities ($v_{H}$).

Spectral linewidths are composed of both thermal and nonthermal(turbulent) components. 
The two components add according to $\sigma_{obs}=\sqrt{\sigma_{th}^{2} + \sigma_{nt}^{2}}$, where
$\sigma_{obs}$, $\sigma_{th}$, and $\sigma_{nt}$ are the observed
linewidth and the thermal and nonthermal components respectively.  
The thermal component arises from the random motions of individual particles within the gas and 
will be different for varying atomic or molecular tracers due to the different particle masses. 
The nonthermal component results from turbulence and bulk motions within the cloud. 
If two species of gas were uniformly mixed throughout a cloud one would expect their spectra 
to share similar non-thermal linewidths. \cite{HINSA1} showed a good correlation between the 
nonthermal linewidths of HINSA and the accompanying $^{13}$CO emission. 
We can use $^{13}$CO to estimate our expected HI linewidths, but first we must disentangle the 
thermal components using information about the gas temperature within the cloud.

Under the assumption that the cold HI is in thermal equilibrium with the surrounding molecular 
gas we can use the kinetic temperature of a molecular tracer to estimate the HI gas temperature 
(T$_{H}$). 
The $J=1 - 0$ transition of $^{12}$CO has been used in the past to estimate gas temperatures 
because it usually appears as optically thick in molecular clouds. 
Equation \ref{maineqn} is valid for molecular emission as well as HI absorption as it describes generic radiative transfer. In the case of $^{12}$CO there is generally no background or continuum emission and so the equation reduces to
\begin{equation}
\label{12COeqn}
T_{A}(v)=T_{^{12}CO}\left( 1 - e^{-\tau_{^{12}CO}(v)}\right) \lc
\end{equation}
where $T_{A}(v)$, $T_{^{12}CO}$, and $\tau_{^{12}CO}(v)$ describe the observed $^{12}CO$ antenna temperature, the $^{12}CO$ gas excitation temperature (which in high density regions can be taken to equal the kinetic temperature), and the optical depth. If the optical depth is taken as $\gg1$ then the equation reduces simply to $T_{A} = T_{^{12}CO}$. 
Thus $^{12}$CO provides us with estimates of the HI gas temperature allowing us to estimate the 
HINSA optical depths, column densities, and thermal and non-thermal linewidths. Some derived $^{12}$CO temperatures for L134 and L1757 are given in Table \ref{HINSAtable2}. Since $^{12}$CO emission has been shown to exist wherever there is $^{13}$CO emission and HINSA we can measure the $^{12}$CO temperature at each position and use it to estimate the $^{13}$CO and HINSA temperatures at each position in the cloud. 
Using these constraints we are left with only one free parameter, the peak optical depth of 
the cold HI gas.

Due to the uncertainties in the assumptions made the constraint parameters cannot be taken as exact values of the HI gas properties. 
Instead they can be used as close guess values to help us constrain the fit. By using $^{13}$CO and 
$^{12}$CO data we are also able to estimate H$_{2}$ column densities using fractional abundances of carbon monoxide that are well--determined in many nearby dense clouds \citep{stahler}.

\subsection{The Fitting Procedure}
\label{fitting}
As explained in section \ref{technique} any narrow HINSA feature will be the dominant feature in the second derivative representation of the observed spectrum.
As such, once guess parameters for the center velocity, line-width and temperature of the HI gas have been 
obtained from molecular tracers the remaining task is to find the value of $\tau_{0}$ for each 
velocity component which minimizes the resulting area under the curve of the second derivative of the recovered background emission spectrum after HINSA is removed in accordance with the discussion in \S \ref{techniqueanalytical}. This is done by first using the trial values of the HINSA linewidth, center velocity, temperature, and optical depth and using those on a particular observed spectrum ($T_{A}(v)$) to obtain the recovered background spectrum without HINSA ($T_{b}(v)$) as described in Equation \ref{maineqn}. According to \S \ref{techniqueanalytical} the HINSA feature will be revealed as a large feature in the second derivative representation of any observed spectrum. To find the best trial $T_{b}(v)$ spectrum the algorithm must search for one whose integraded intensity in the second derivative representation is minimized. The second derivative of the recovered spectrum ($T_{b}(v)$) is calculated numerically through traditional methods. Since the second derivative representation will have negative values as well as positive ones, it is squared prior to integration. We thus minimize the value of $I$ as described by the function
\begin{equation}
\label{tominimize}
I = \int \left(\frac{d^{2}T_{b}(v)}{dv^{2}}\right)^{2} dv \lp
\end{equation} 
which is simply the squared integrated intensity of the second derivative of the background spectrum ($T_b$) that is recovered after trial values of the HINSA optical depths and other parameters are removed from the observed spectrum. Finally, determining the smallest value of $I$ will give us the best-fit values of the HINSA parameters. 
It is possible to use a more sophisticated minimizing function, but the HINSA feature is so 
dominant in the second derivative representation that this has proven not to be necessary.

Previous studies have shown that the correlation between HI and $^{13}$CO center velocities 
is strong \citep{HINSA1}. 
Hence, the $v_{H}$ parameter is held fixed for each velocity component. 
The $T_{F}$ estimates obtained from $^{12}$CO do not carry as much confidence 
(as discussed in the next section), however the temperature is still kept as a fixed parameter 
during our fit. 
For optically thin spectral lines such as HINSA the temperature cannot sufficiently constrain 
the shape of the absorption feature to be allowed as a free parameter since a wide range of combinations of temperatures and optical depths would be able to fit the data. Thus the gas temperature must be specified in any fits.

The linewidths obtained from $^{13}$CO observations can only be applied to the HI gas 
under the assumption that the two species are coextensive throughout the cloud. 
This assumption, while strongly suggested by the previously--referenced studies, is clearly not exact. 
Hence the molecular linewidths for each velocity component are used as guess values for the HI fit. 
The HI linewidths are allowed to vary in value during the fit near the initial guess values.

In practice, $\tau_{0}$ is left as a free parameter, and $\sigma_{H}$
as a constrained parameter for each component. For every trial value
of $\tau_{0}$ a T$_{b}$ spectrum is constructed and its second
derivative is checked for the fitting parameter as specified by
equation \ref{tominimize}. All velocity components along the same line of sight must be fitted 
simultaneously to account for obscuration due to overlapping. 
There are a variety of fitting algorithms that can be used for the final fitting. 
In the present case a variation of the Levenberg--Marquardt method was implemented. 
In the interest of clarity, we give in appendix \ref{processAppendix} a step-by-step description of the entire process including the determination of molecular parameters and HINSA extraction.

\section{Coming to Terms with Variable Beam Sizes and Unresolved Sources}
\label{beamsizes}
Our technique combines observations from different telescopes with different frequencies and beam sizes. 
The HI data presented in this paper was obtained using the 100 m Greenbank Telescope (GBT) having a 9' beam size, while the $^{12}$CO and $^{13}$CO data were obtained using the Five College Radio Observatory (FCRAO) 14 m telescope having a 45'' beam size. 
Combining these data sets is non-trivial. 
Convolving the CO data directly to a 9' beam has the effect of smoothing out a great deal of structure, kinematically as well as spatially, which is neither necessary nor desirable.

By fitting the molecular data prior to convolution we preserve information regarding individual velocity components which may otherwise disappear or be distorted in an HI beam. 
It is not necessary for all molecular components to have accompanying HINSA features. 
Maintaining this information is crucial to studying the comparative composition of different 
velocity components within a cloud as well as properly determining
their HI/H$_{2}$ ratios. 
High-resolution molecular data should only be convolved to a beam size which results in
sufficient signal--to--noise to produce good fits to the individual emission components. 
Following this argument and a series of numerical experiments, our FCRAO spectra were convolved to a 2 arcminute beam size. 
Parameters including center velocity, linewidth, temperature, and optical depth for each molecular emission velocity component are derived using conventional fitting techniques. 
Corresponding guess parameters for the HI gas are then constructed by convolving the molecular fit parameters to the HI beam. 
This approach allows us to measure the HINSA column densities for each individual molecular component, as well as to utilize the high spatial resolution information on the molecular constituents of the cloud.

Many sources are unresolved when observed with available single-dish HI beams. 
Previous studies (\cite{HINSA1}, and \cite{HINSA2}) have shown strong spatial and linewidth correlations between HINSA and $^{13}$CO. 
With this information we may make the assumption that HINSA and $^{13}$CO are well mixed 
throughout a cloud. 
This affords us the opportunity to make estimates of the HI beam filling factors by 
assuming that HINSA is present wherever significant $^{13}$CO emission is observed. 
This assumption, while not precise, is useful. 
In the case of resolved sources the beam filling factor plays a minimal role. 
In unresolved sources, or the edges of larger clouds, the beam filling factor as implemented by the present technique, does not alter the calculated HI/H$_{2}$ ratio, but only alters the total column densities of both species.

\section{Example Solutions Using Simulated Data}
\label{examples}

A controlled way to assess the validity of our technique is to apply it to synthetic, yet realistic data and compare the 
derived results with the known input values. 
Simulated data does not allow us to test the assumptions on the relations between molecular and atomic gas that are used by this technique. 
It is possible however to assume those relations to be correct and to test the validity of the HINSA 
fitting technique by itself. 
Further it is possible to observe the effects on the HINSA fit of having incorrectly determined the 
fitting parameters based on molecular data.

\subsection{A Simple Case}
\begin{figure}
\epsscale{1.0}
\plotone{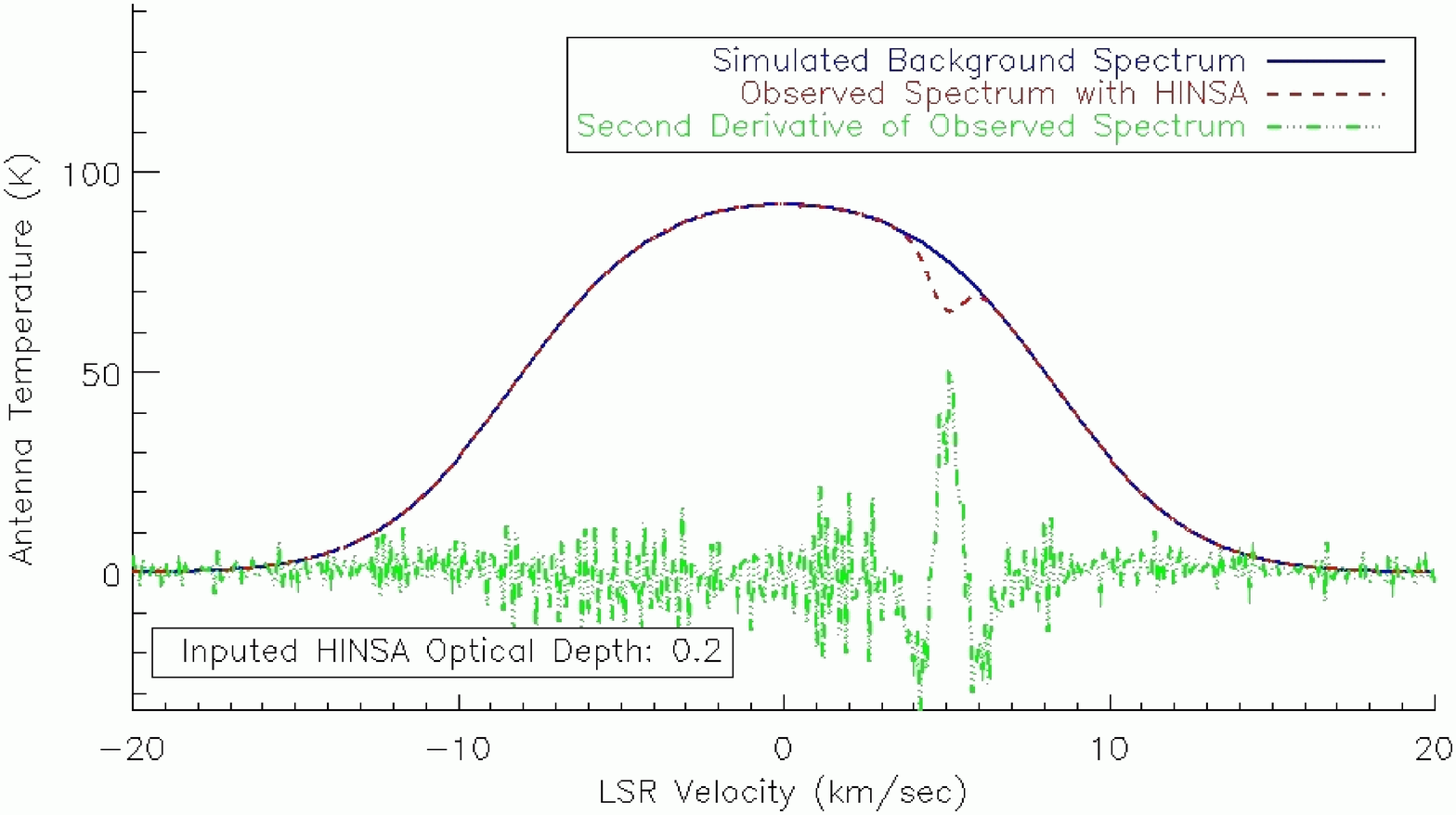}
\caption{ 
An idealized HINSA spectrum (red) from simulated data. 
The blue line represents the background HI emission spectrum, while the green line represents the second velocity derivative of the observed HINSA spectrum. The spectra include typical radiometric noise. The HINSA feature has a peak optical depth of 0.2 and gas temperature of 10K. 
\label{simple1}
}
\end{figure}

\begin{figure}
\plotone{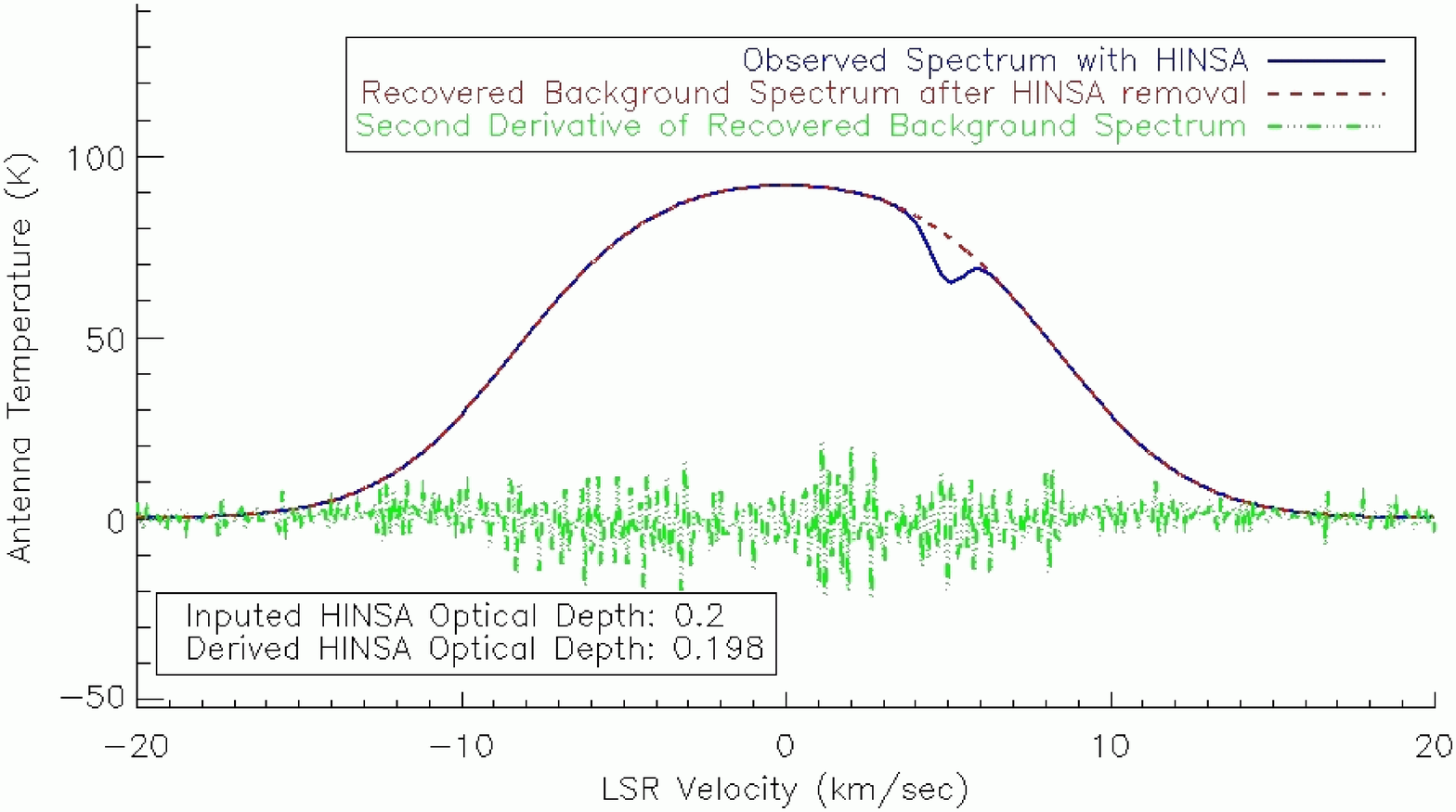}
\caption{
The original observed spectrum (blue) along with the derived background spectrum after HINSA removal (red) using the same simulated data as in Figure \ref{simple1}. 
The recovered background spectrum, and the calculated optical depth of 0.198 are nearly identical to the input parameters. 
The second derivative representation of the recovered background spectrum (green) shows no signs of the HINSA feature.  
\label{right1}
}
\end{figure}

The simplest case consists of a single emission component obscured by a single HINSA absorption feature as in Figure \ref{simple1}. 
As expected, HINSA is the dominant feature in the second derivative representation. 
For the simulated run, the center velocity and temperature of the cold
HI gas are provided as precisely known quantities. 
The linewidth, however, is given only as a guess parameter. 
The task then is to search for those values of the linewidth and the peak optical depth which minimize the squared sum of the second derivative background spectrum after HINSA is removed. 
Figure \ref{right1} shows the original observed spectrum as well as the recovered background spectrum after fitting. 
The resulting computed optical depth of 0.198 compares favorably to the original inputed value of 0.2. 
It is apparent from the second derivative representation of the recovered spectrum that the HINSA which was prominent in the original spectrum (Figure \ref{simple1}) has been effectively removed.

\subsection{Effects of Faulty Parameter Determination from Molecular Data}
\label{faulty}

It is clear from the previous simulation that in an idealized case where the properties (center velocity, 
linewidth, and temperature) of the HI gas are well known, the HINSA optical depths can readily be determined to a high degree of accuracy. 
However in practice, the properties of the HI gas are not known but are instead surmised with some uncertainty from observations of molecular tracers. 
Hence it is necessary to examine the impact of incorrectly deriving the HI properties from molecular data.
One concern is the gas temperature, which we obtain from $^{12}$CO.    
It is reasonable to assume that the HI excitation temperature is equal to the kinetic temperature of the gas.
However, temperature gradients may exist resulting in the average temperature appropriate for analyzing the HINSA being different from that determined from the optically thick common isotopologue of carbon monoxide. 
Since HINSA features are typically optically thin, an incorrect estimate of the HI gas temperature would not strongly alter the shape of the fitted HINSA, but would only affect the derived optical depths and column densities. 
Thus, incorrect temperature estimates would be difficult to detect through fitting alone.

\begin{figure}
\plotone{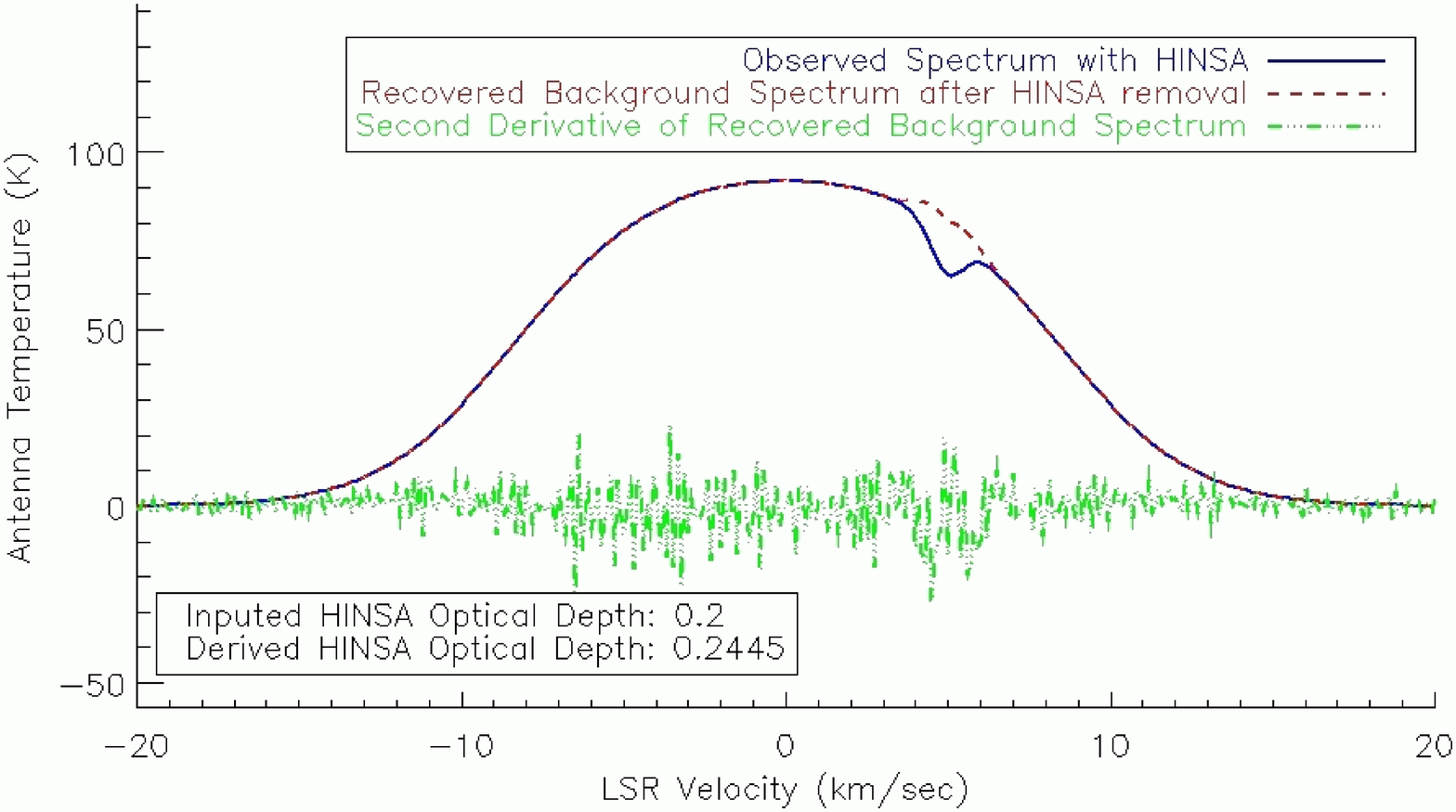}
\caption{
Using a HINSA linewidth 20\% greater than the correct value when fitting the same simulated HINSA spectrum as in Figure \ref{simple1} (blue) produces a distorted recovered background spectrum (red).  The second derivative representation has a clear signature at the velocity of the HINSA absorption feature observed, and both its apearance and that of the the recovered background spectrum are unlikely to be representative of a realistic system.\label{wrong2}
}
\end{figure}

\begin{figure}
\plotone{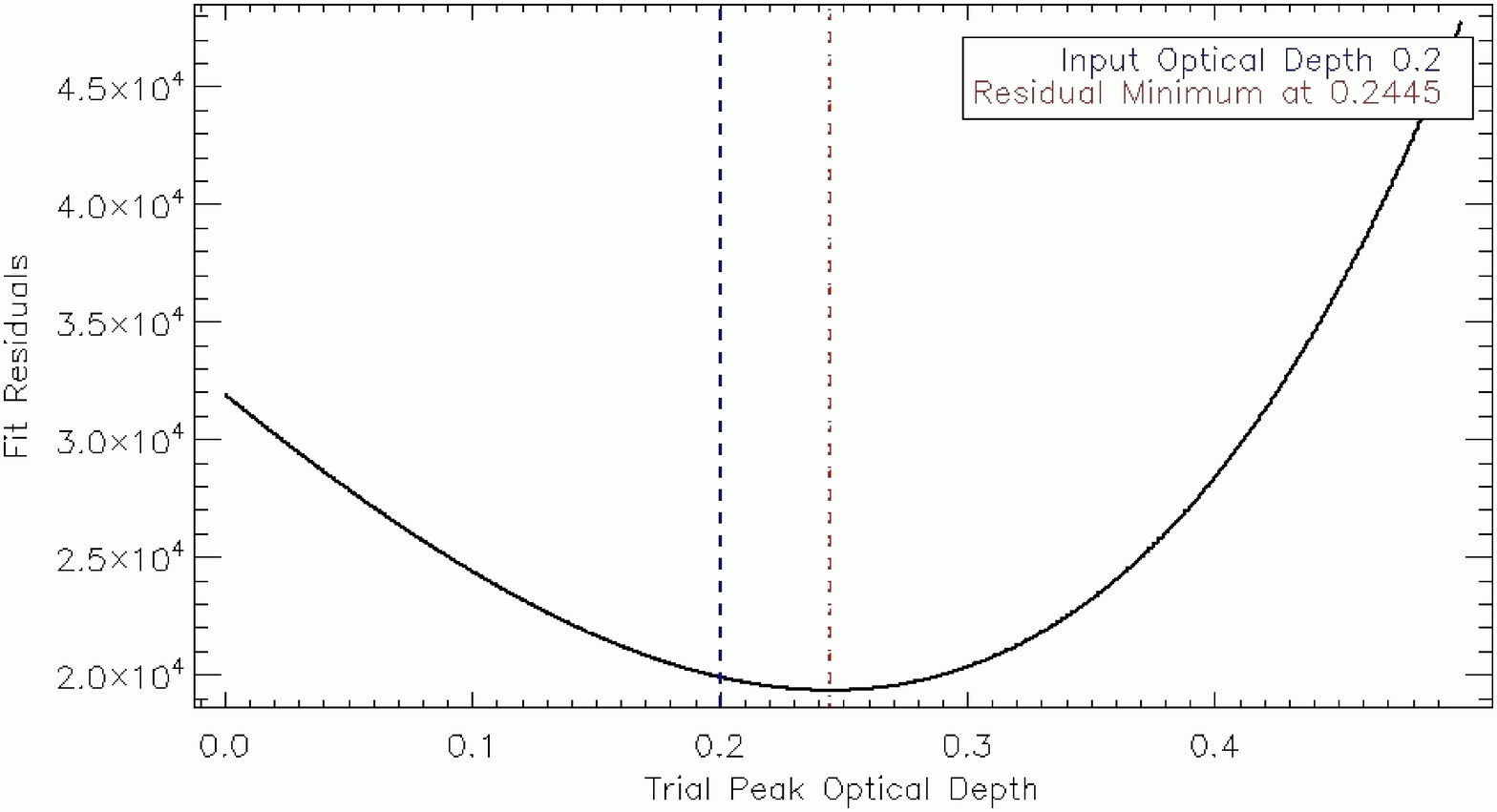}
\caption{The squared sum residuals of the second derivative of the recovered background spectra 
for different trial values of the optical depth using the same data, and incorrect linewidth as in Figure \ref{wrong2}. This function $I$ represents the quantity to be minimized in our fitting procedure described in \S \ref{fitting}.
The minimum value of $I$ is considered to be the best fit, and in this case corresponds to an optical depth of 0.245, compared to the correct optical depth of 0.20. Realistic simulated statistical noise is included in this plot, and while the range in optical depth corresponding to a specified increase in the residuals may be significant, there are no localized minima to confuse the results.
\label{wrong1}
}
\end{figure}

Incorrect estimates of the HI linewidths based on $^{13}$CO data are plausible since the relation between them is only approximate. 
Here we investigate the effect of such an error by forcing our fitting routine to find the best 
fit HINSA optical depth while being constrained to use an incorrect value for the HI linewidth. 
\cite{HINSA1} showed a variance of roughly 20\% in the observed relation between HINSA and $^{13}$CO 
nonthermal linewidths. 
Figure \ref{wrong2} shows the resulting fit when forced to utilize a HINSA linewidth 20\% larger 
than the actual value of the same simulated data as in the previous example. 
It is no longer possible for the algorithm to produce a smooth recovered background spectrum, regardless of the optical depth assumed. 
This is a key indicator of an error in the input information. 
Figure \ref{wrong1} shows that the new best fit HINSA optical depth is 0.245 rather than 0.20.  
Similarly inaccurate solutions result when a faulty center velocity is used.

\subsection{Complex Examples}
\label{complex}

It is often the case that two emission components may overlap at the velocities at which HINSA may 
be expected, thus obscuring the absorption feature and increasing the uncertainty in the analysis. 
In such cases, previous techniques have had great difficulty in determining whether any HINSA is present at all. 
Our technique attempts to recover the smoothest, most gaussian-like background spectrum while being constrained 
by the HI gas parameters derived from molecular observations. 
It is unlikely for emission components to appear over a HINSA features in such a way as to completely obscure the absorption.

\begin{figure}
\plotone{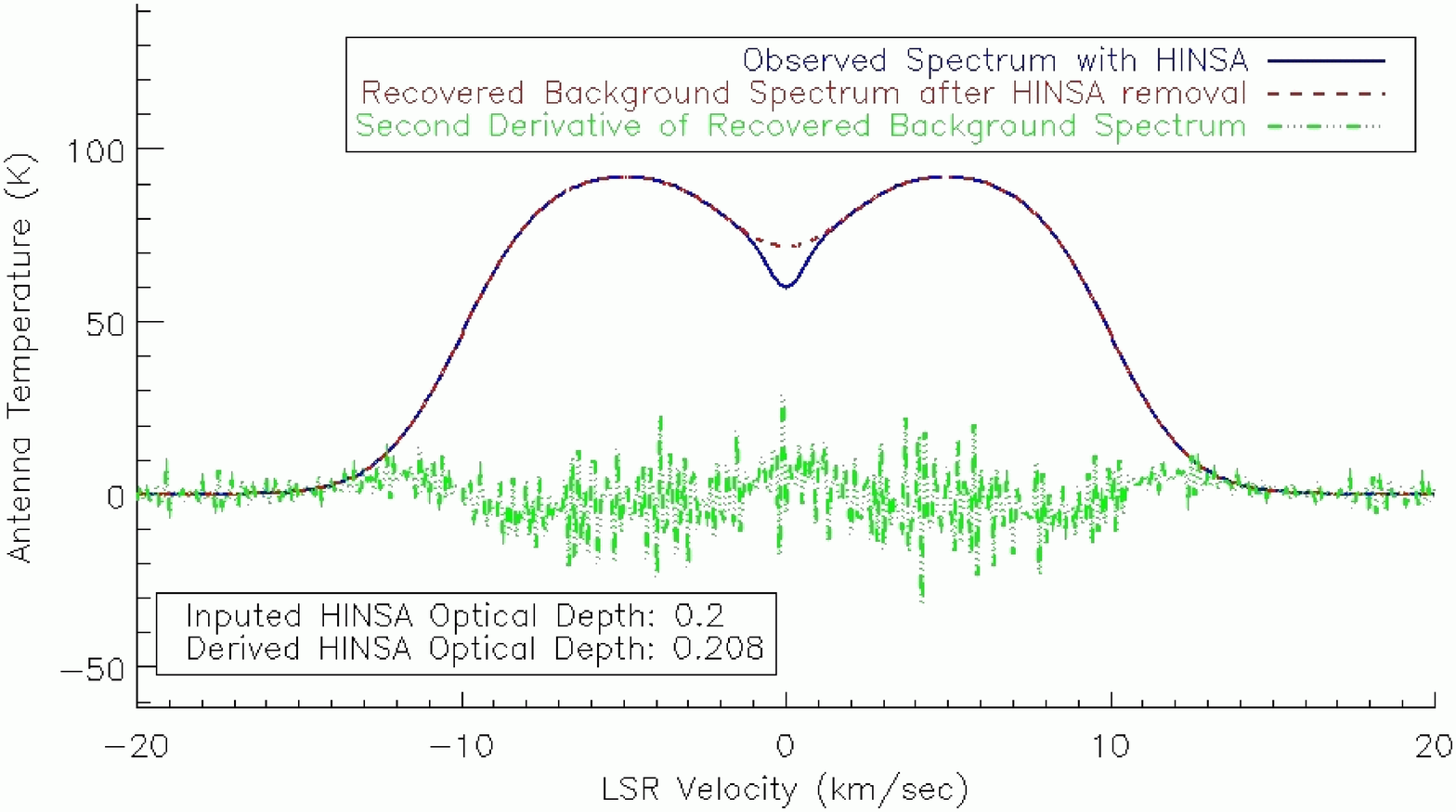}
\caption{
Two relatively broad identical emission components combined with a HINSA feature located midway between them yield the simulated observed spectrum (blue). 
The recovered background spectrum (red) and accompanying second derivative representation (green) illustrate 
the results after HINSA extraction. 
The derived optical depth is 0.208, which compares very well with the input optical depth of 0.20.
\label{complex1}
}
\end{figure}

\begin{figure}
\epsscale{0.8}
\plotone{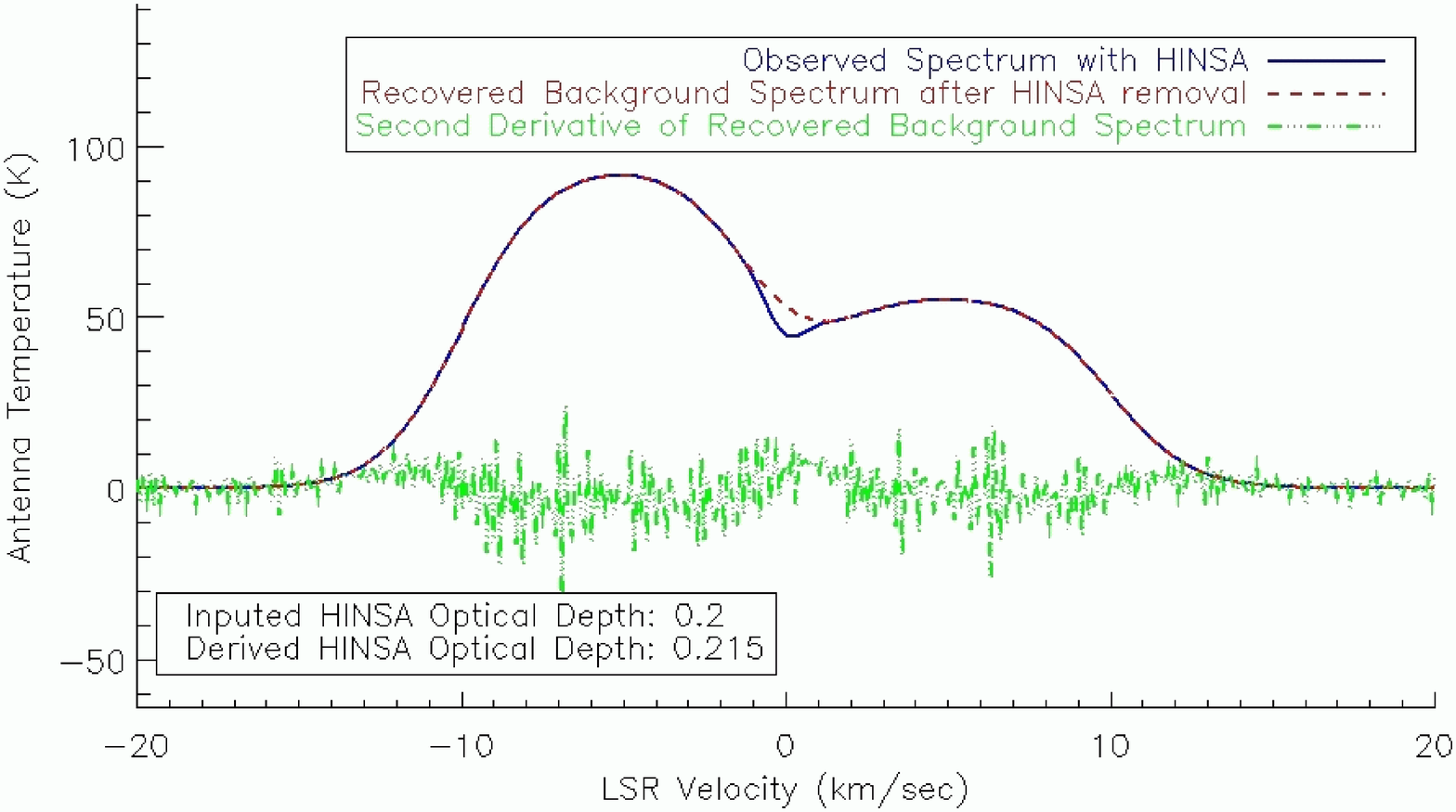}
\plotone{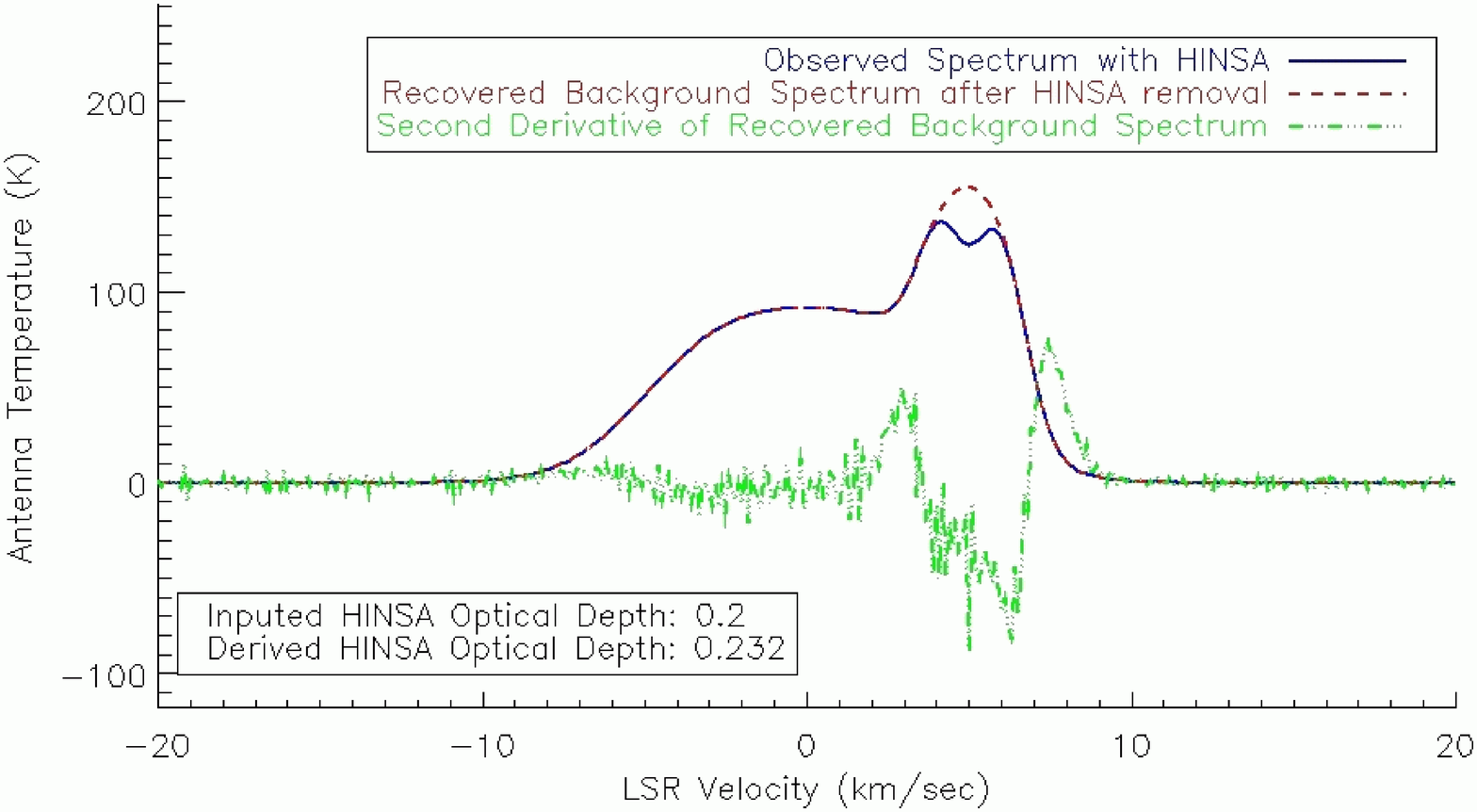}
\caption{
Upper: Two unequal emission components together with a 0.20 peak optical depth HINSA feature result in the simulated observed spectrum (blue). 
The recovered HINSA optical depth is 0.215, and the HINSA extraction results in the recovered background spectrum (red), which is nearly identical to the input background spectrum. 
Lower: A similar case where the HINSA feature has been placed to coincide with a narrow emission component whose 
shape is slightly distorted by the presence of another emission component. 
The derived HINSA optical depth of 0.232 is still close to the value 0.20 that was input to the calculation. 
The second derivative representation of the recovered background spectra (green) shows that there is still 
significant structure left after HINSA extraction. 
This structure is the result of the interaction between two emission components. 
\label{complex2}
}
\end{figure}

As an example, Figure \ref{complex1} shows that our technique is able to accurately extract HINSA features which on visual inspection might be ambiguous. 
In this simulation, the HINSA feature with an optical depth of 0.20 was placed in between two identical 
emission components. 
The recovered HINSA optical depth had a value of 0.208 which is well within the error limits of other 
uncertainties as discussed in the section. 
Figure \ref{complex2} shows further examples with more complex geometry in which our technique is 
still able to pick out HINSA features with good precision.

\subsection{Distorted Emission Components}
\label{distorted}

\begin{figure}
\plotone{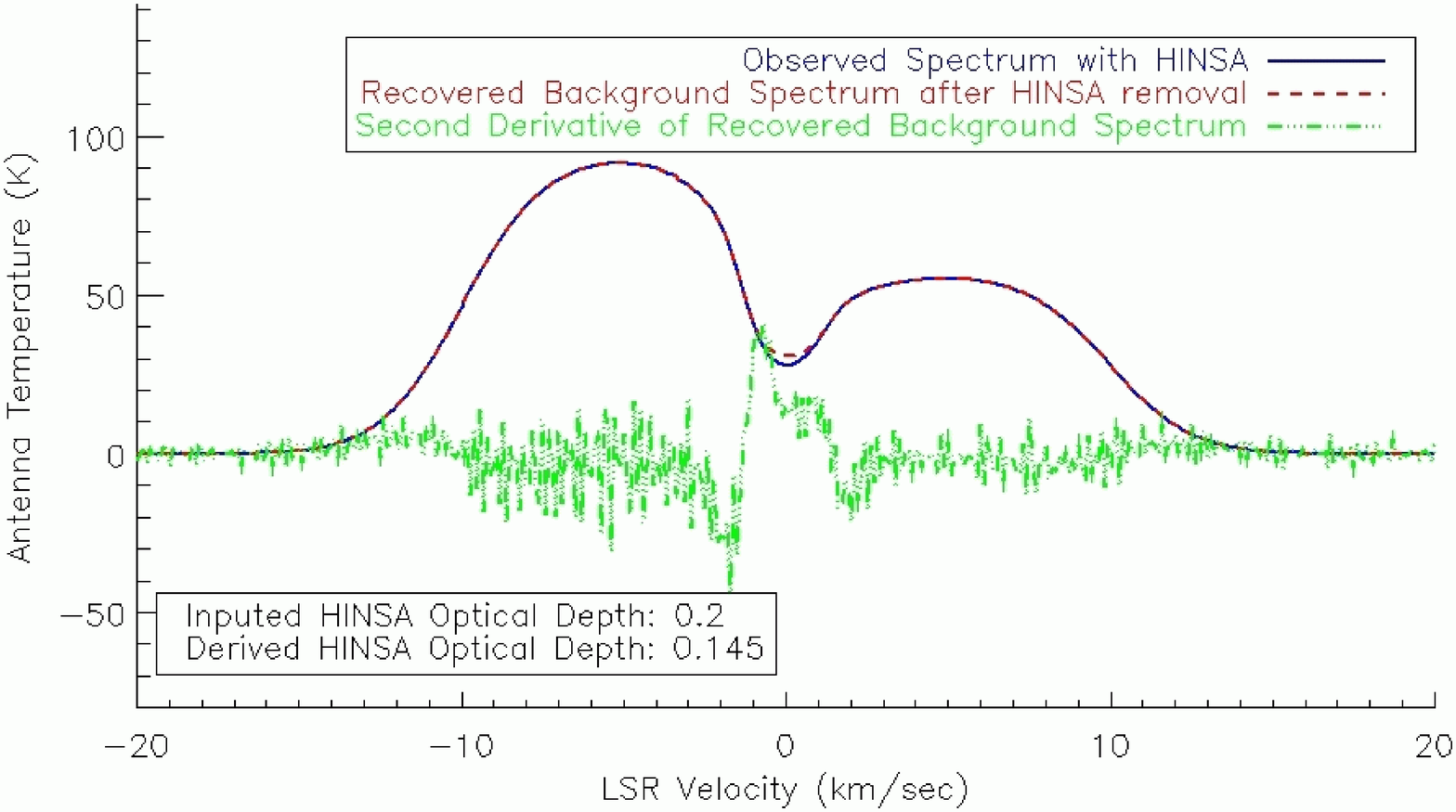}
\plotone{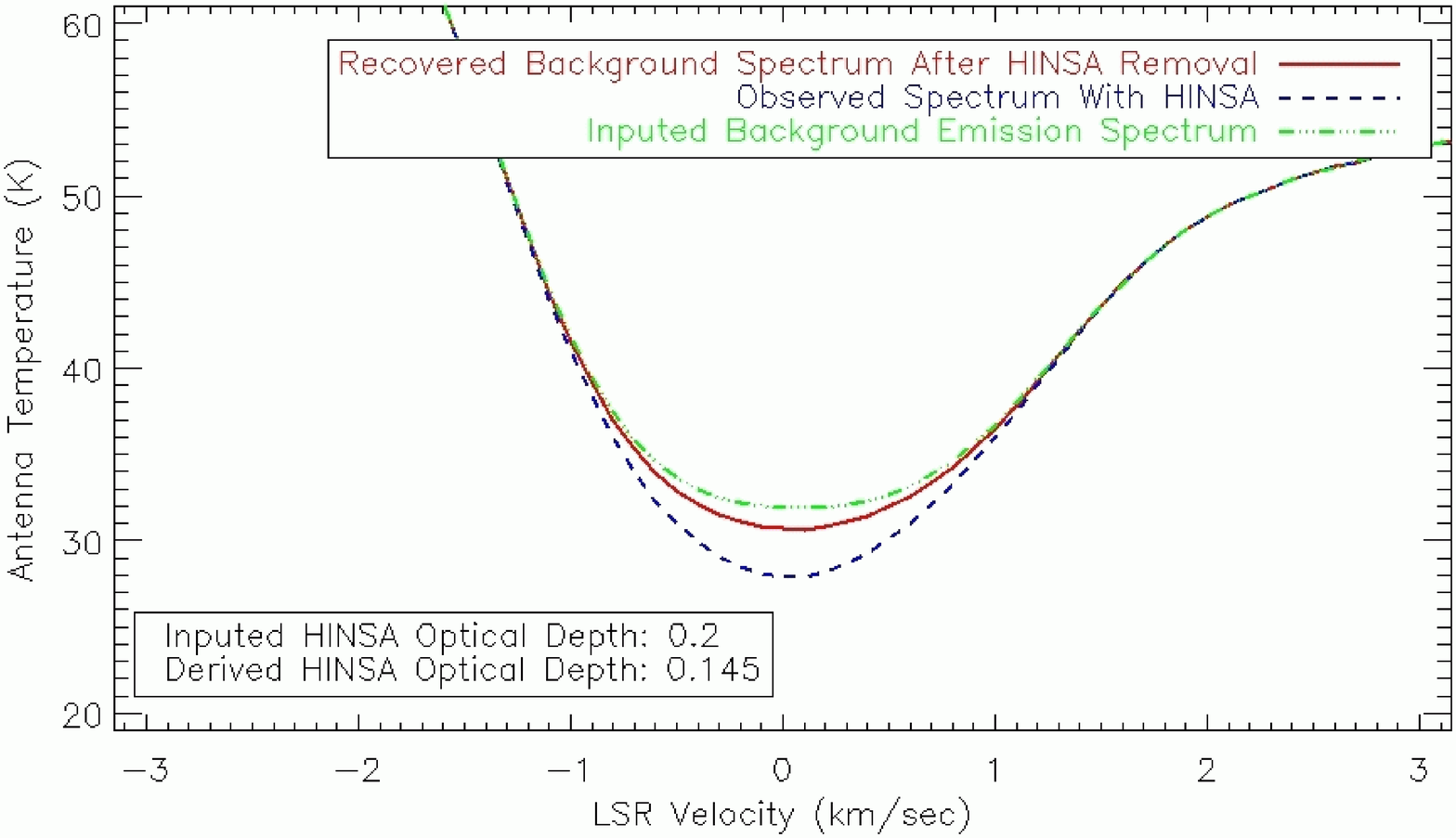}
\caption{
Upper: For comparison with Figure \ref{complex2}, a third emission component with brightness temperature of 30 K and center velocity of 0 \kms\ has been introduced to distort the shape of the emission spectrum at the location of the simulated HINSA feature. 
In this instance the derived HINSA optical depth of 0.145 differs considerably from the inputed value of 0.2. 
Lower: An expanded view of the region of interest showing the recovered and inputed background emission spectra.
\label{complex3}
}
\end{figure}

\begin{figure}
\plotone{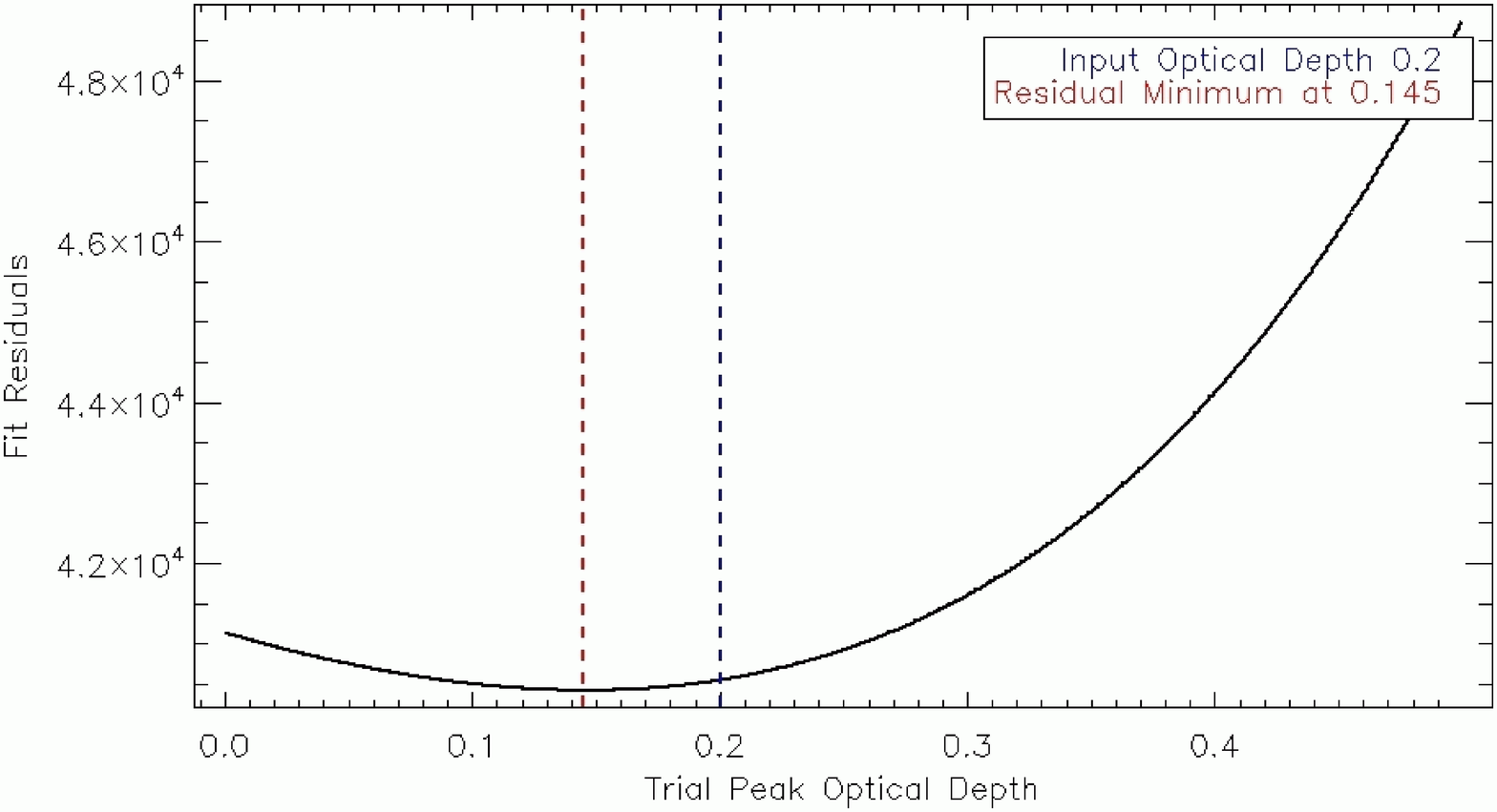}
\caption{
A similar residuals plot as in Figure \ref{wrong1} as applied to the solution in Figure \ref{complex3}. The behavior of the residuals is the same here even though the error in the solution is clearly evident ($\tau$ = 0.15 compared to 0.20). This demonstrates the difficulty in identifying incorrect fits in cases such as in Figure \ref{complex3}.
\label{residual3}
}
\end{figure}

The technique we present here is based on the assumption that the background emission is composed primarily of a 
superposition of gaussian--like components.  
Significant deviations from this assumption can lead to incorrect results. 
Figure \ref{complex3} shows the case in which the background spectrum is complex, due to the superposition of overlapping components.  
The presence of a background component of width only modestly greater than that of the HINSA feature, in combination with the broader emission feature, results in a noticeable error in the derived HINSA optical depth. 
On casual inspection, there may be little indication that the solution is problematic (as shown in the lower panel of Figure \ref{residual3}).
When dealing with real data, where the actual background spectrum is not known beforehand, there is no simple 
way to determine when such distortions may occur, or what their effect might be on our derived HINSA optical depths. 
Though instances such as this are likely to be rare, it is worthwhile to be especially wary of HINSA measurements 
when the absorption feature seems to lie in the trough between two emission components.

\subsection{Confidence in Derived HINSA Optical Depths from Simulated Data}
\label{confidence}

It may be tempting to take Figure \ref{wrong1} and use standard methods for determining the uncertainties in least-squares linear fitting problems to determine the uncertainties in our estimates of $\tau$ based on the width of the trough of the fit residuals. But our problem does not strictly allow such analyses. There are two components that contribute uncertainty to our estimates of $\tau$. The first is radiometric noise. The simulated data in this section all include realistic simulated noise for observations made with a large single-dish telescope such as Arecibo or the GBT. This source of uncertainty is in fact linear in nature and could be treated using traditional least-squares linear fitting analyses. However, as HI spectra are typically bright and easily obtainable with only a few seconds of integration time this source of uncertainty is insignificant compared to the other uncertainties. If we were to look directly at Figure \ref{wrong1}, then the uncertainties due to radiometric noise would appear as small, narrow localized minima. However, in the case of this and all our simulations this contribution is smaller than even the thickness of the line in the plot.

The dominant source of uncertainty which produces the trough is non-linear and somewhat systematic. The fundamental principle on which our technique depends, as discussed in \S \ref{techniqueanalytical}, requires that our observed spectra resemble the superposition of multiple (positive and negative) gaussians. According to the radiative transfer equation however this is not strictly the case. When multiple emission and absorption components are superimposed they attenuate one another in a non-linear fashion. Hence our observed spectra will never strictly represent gaussians, but will only resemble them. This introduces a somewhat systematic, ''nonlinearity'' error which is much greater than that introduced by radiometric noise.

As an example, if we were to re-run the simulation from Figure \ref{wrong1} numerous times with all the same parameters for emission and absorption components but different radiometric noise (of the same average amplitude), the measured optical depth would always be very close to 0.245, the only variation in the results would be due to radiometric noise which in these realistic cases is insignificant. This represents a 22.5\% error (not an uncertainty) relative to the original inputed value of 0.20 which is produced not by random local minima which would confuse the results, but by a systematic shifting of the entire residuals trough. This systematic error is produced when several emission and absorption components are superimposed in such a way that the final spectra no longer resemble simple superpositions of gaussians. Throughout this section we have shown successively more complex superpositions of this nature in an attempt to test the limits of this systematic error. We have found that for the vast majority of observations likely to be encountered, this error will be only a few percent. In an attempt to find the limits, we constructed the most complex superpositions that would likely be observed in real data and found that in the very worst cases the systematic error reached only about 50\%. Figures \ref{complex1}, \ref{complex2}, and \ref{complex3} represent some of those complex cases. For the majority of observations, this nonlinearity error will be minor in comparison to other sources of error due to uncertainties in physical conditions in the cloud. In the most extreme cases, the nonlinearity error will be on a par with, but not dominant compared to these other sources of uncertainty, which are discussed in the following section.

These simulations cannot directly test the validity of our assumptions regarding the relations between atomic and molecular gas properties as those are inherently non--statistical uncertainties whose properties are currently not well known. However, the simulations do show that the solutions are modestly sensitive to any such errors in that errors in our $^{13}CO$--based estimes of either the HI linewidths or center velocities neccessarily produce visible artifacts in the recovered background spectra after HINSA extraction, as shown in Figure \ref{wrong2}. This allows us to identify faulty fits, where we can then go back to the molecular data and construct better templates for the HINSA extraction. Hence it is important to understand all the possible sources of uncertainty in our estimates of the HINSA properties based on molecular templates as discussed in the following section.

\section{Limitations: Sources of Ambiguity and Uncertainty}
\label{ambiguity}
While it has been shown in the previous section that our technique is able to produce good fits to the HI data when the HI gas properties are correctly derived, there are several factors which limit the accuracy of our technique. 
These sources of error are non--statistical in nature in that they arise from our reliance on the association between HI and molecular tracers. 
As such, they make the placement of simple error bars on derived HI optical depths impossible. While it is certainly possible to place uncertainty estimates numerically through standard least-squares fitting methods, and compound those with estimates of the uncertainties due to measurements of the molecular parameters and their relation to the HI gas, any such numbers would be meaningless and misleading. The best way to estimate the uncertainties in our results is to examine the spread in the results in Figures \ref{HINSAplots} and \ref{HINSAplots2} and assume that in an ideal case each component would be represented by a single line (of certain shape) which would be governed by the physical processes within the cloud such as the HI to H$_2$ conversion process. This would give us an upper limit on our uncertainty in deriving the HI/H$_2$ ratio. However, the present models of cloud evolution do not yet offer us sufficient confidence to make such estimates with certainty. More detailed observations and modeling of molecular cloud processes are necessary before such estimates can be made with confidence. It is our hope that the technique described in this paper will help in this regard. This development is likely to be an iterative process. It is important to note that the sources of uncertainty in previous techniques, while inherently somewhat different, are not smaller nor better determined than those in the method presented here. It is, however, of critical importance to quantify and to understand each of these sources of uncertainty in order to understand the level of confidence in the results obtained using this new technique.

\subsection{Gas Temperature}
\label{temperature}

Knowledge of the HI gas temperature is necessary for estimating the non-thermal linewidths of $^{13}$CO as well as the observed linewidths of the HI gas, the measurement of the HI optical depth, and column densities both for HI and molecular tracers. 
As such, accurate temperature estimates are critical.  
At the present time the most practical method for estimating the HI gas
temperature in these dark cloud cores is through obtaining the $^{12}$CO temperature. 
Typically, $^{12}$CO emission is optically thick, and thus if large scale velocity gradients are absent, as is generally the case in dark clouds, this measurement represents only the surface temperature of the cloud and does not necessarily reflect the average HI gas temperature in the interior of the cloud. 
The cores of these clouds are shielded from the external UV radiation field and thus are likely to be at a lower temperature than the surface. 
While this bias introduces overestimates of the HI gas temperature by only a few degrees, typical core temperatures are 
on the order of 10K thus the effect is significant.
It may be possible to use alternate means to estimate the temperatures of other molecular tracers which share 
similar distributions as HI within a cloud. 
One promising candidate is the inversion transitions of ammonia.
Detailed comparison of this or other tracers with \tw\ would require a better understanding of the spatial distribution 
of cold HI necessitating interferometric maps of HINSA features.

\subsection{Radiometric Noise and Baselines}
\label{noise}

21cm spectra with high signal to noise ratios are relatively easy to obtain with short integration times. 
Radiometric noise should thus not significantly affect HINSA measurements. 
Obtaining good baselines on the HI spectra is important in that any artificial slopes can significantly 
alter the derived optical depths. 
Since HI emission is highly variable spatially, ON-OFF observations are impractical.
Total power ON measurents are ideal if system stability permits this approach to be used, as reported by \citep{HINSA1}. 
Frequency-switched observations can yield sufficiently good baselines, but only if system bandpass and calibration are sufficiently good. 
A very successful combination of these techniques is available with the recently installed Arecibo L-Band Feed Array (ALFA) in conjunction with the Galactic-ALFA (GALFA) spectrometer at Arecibo.
Least-Squares Frequency Switching (LSFS) allows the RF and IF bandpasses to be separately determined so that total power observations produce sufficiently good baselines for HINSA observations (Krco et al. 2008, in preparation).

\subsection{Choice of Molecular Tracer}
\label{choice}
\cite{HINSA1} showed good correlations between the central velocities, nonthermal linewidths and spatial distributions of HINSA and both $^{13}$CO and C$^{18}$O. 
In many cases, $^{13}$CO shows a better spatial correlation, but C$^{18}$O shows a better linewidth correlation. 
This would not be expected if the relative abundances of HI, $^{13}$CO, and  C$^{18}$O were constant 
throughout the cloud. 
The variations may be a reflection of the higher critical density of C$^{18}$O, the density dependence of the 
HI/H$_{2}$ conversion process, and possibly CO depletion onto dust grains at high densities. 
Thus the choice of which molecule makes the best template for the HINSA linewidth is complex. 
Based on experience,  using $^{13}$CO fits provides somewhat better results.
As a practical matter, the  $^{13}$CO lines are much brighter and easier to observe than those of C$^{18}$O. 
As a result, we have selected $^{13}$CO as the template molecule for HINSA extraction.

\subsection{Distances and Foreground Gas}
\label{distances}

The presence of intervening HI emission sources between the observer and the HINSA source will tend to wash out the 
absorption and introduce underestimates of the HINSA optical depth. 
\cite{vanderWerf} introduced an adjustment factor to compensate for this effect, however its precise value is poorly known.
This approach was followed by \cite{HINSA1}, but these authors found it to be
relatively unimportant for the relatively nearby clouds in the Taurus
Molecular Cloud Complex. A more detailed study of the relation between HINSA features and distance is required when dealing with more distant sources.

\subsection{Component Order}
Several velocity components are often observed in molecular clouds. 
It is assumed that there are correspondingly multiple HINSA components. 
Equation \ref{maineqn} indicates that when multiple overlapping HINSA components are present along the same line of sight with appreciable optical depths, there is significance in which component is placed in the background, and which is in the foreground. 
There is no simple way to determine the sequence of multiple components along the same line of sight. 
Since the typical optical depths for HINSA are on the order of ~0.1, the resulting ambiguity for the 
HI column densities is below ~10\%, smaller than the other sources of error in our technique.

\subsection{Molecular Emission Fits and the Uniqueness Problem}
The center velocities of an individual molecular component may vary significantly within a cloud. 
It is often the case that two components may be widely separated in velocity at one point in the cloud while being 
nearly indistinguishable in another region. 
This poses a problem when trying to make measurements on individual components and is especially pronounced when working with HI spectra with high thermal linewidths. 
Further, it is difficult to predict how a specific implementation of the technique proposed here might react 
when working with such spectra. 
When working with a sufficiently small number of spectra it is possible to recognize and dismiss the affected spectra by eye. 
However, in larger regions, where there may be many thousands of spectra, and where fully automated fitting algorithms must be used, the problem becomes more pronounced. 
Thus we have developed a method for partly alleviating the uniqueness problem and allowing for easier identification of overlapping emission spectra. 
This is a problem most often avoided in many projects involving complex emission spectra since usually only the total integrated intensity is used. 
For the technique presented here however, the individual properties of each emission component are necessary.

Molecular emission spectra are typically characterized by (overlapping) gaussian components. 
It is a property of gaussians that taking any gaussian function to the fourth power effectively reduces its linewidth by a factor of 2. 
Therefore, by taking all our molecular spectra to the fourth power prior to fitting, the individual emission components become significantly more apparent. 
However, doing this also greatly magnifies the noise in the spectra. 
Thus it is necessary to perform some filtering by removing the high-frequency Fourier components of each spectrum. 
The magnification and filtering processes can significantly alter the amplitude, shape, and linewidths of the final spectra. 
However, the final center velocities remain the same. 
Thus, the filtered spectra may be used to identify the number and center velocity of the emission components within a spectrum. 
Subsequently, the known velocities can be used to constrain the fitting of the original data. 
While not yielding a unique solution, this method does allow more accurate fitting of molecular emission components using automated algorithms and dealing with large spectral data sets.

\section{Summary and Future Work} 
\label{summary}
The significance of accurately measuring the HI content of dark clouds and star forming regions has long been recognized. 
HINSA features have been shown as a promising method to achieve this goal, but the difficulty in confidently disentangling HINSA from the galactic background emission has limited research efforts in this field. 
The technique described here builds upon previous work to provide new opportunities. 
By utilizing a second derivative representation in which HINSA becomes the dominant feature in the spectrum, 
and using information from associated molecular tracers to constrain our fits, we are able to obtain HI column 
densities with greater confidence than previously possible.
This technique enables us to study individual velocity components within a cloud. 
Several uncertainties and sources of error still remain, most notably the errors inherent in temperature determination 
through $^{12}$CO, and the precise relations between the properties of cold HI and molecular tracers such as $^{13}$CO. 
As shown in \S \ref{examples} the purely statistical errors derived from our constructed data were only a few percent in simple cases which represent the majority of observed spectra.  The nonlinearity error may be as large as 50 percent in the extreme cases.

While the scope of this paper is limited to a demonstration of the new technique, the improved confidence in the 
results have the potential to yield significant scientific advances in the field of molecular cloud and star 
formation studies which are deferred to a subsequent publication. 
The two astronomical sources discussed briefly here for demonstration purposes are part of a much larger survey of over 30 dense cloud cores. 
In conjunction with numerical modeling these data provide us with the chemical ages of the clouds and individual 
components therein, thus providing a significant constraint on theoretical models of star formation. 
Large maps have been collected with the Arecibo L-Band Feed Array (ALFA) of the Taurus and Perseus star forming regions. 
These regions show plentiful HINSA features whose analysis may shed light on the dynamics of such complexes including the processes which may have triggered their formation.

\section{Acknowledgments}

This research was supported in part by NSF Grants AST 0404770 and AST 0407019, by the NRAO GBT Student Support Program, and by the Jet Propulsion Laboratory, California Institute of Technology.
The data used in this paper were obtained with the Five College Radio Observatory, which is operated with support from the National Science Foundation and with permission of the Metropolitan District Commission, and with the Greenbank Telescope (GBT) of the National Radio Astronomy Observatory. 
We thank Mark Heyer and Jay Lockman for assistance in carrying out the observations, and Jim Cordes, Carl Heiles, and Chris Salter for useful discussions.

\newpage
\begin{appendix}
\section{Description of the Procedure}
\label{processAppendix}

For definitiveness, we here describe the process of obtaining molecular parameters to be used 
as a template as well as the actual HINSA extraction in a step--by--step fashion. 
We model this description based on our specific observations and available data although the technique can be
easily adapted to other circumstances. 
We obtained $^{12}$CO and $^{13}$CO data with the 14 m Five College Radio Observatory having a 45'' to 50'' FWHM
beam size and the HI data were obtained using the 100 m Green Bank Telescope having a 9' FWHM beam size.

\begin{enumerate}
\item Reduction and Regridding: It is assumed that all data have been properly reduced, calibrated, and regridded.

\item Convolution: Initially the $^{12}$CO and $^{13}$CO spectra are convolved to a 2' FWHM beam size in order to improve the signal to noise ratio. A gaussian convolving beam shape is used.

\item CO Fitting (first pass): While assuming a particular gas temperature (10K), emission functions are fitted to all $^{13}$CO spectra to obtain their center velocities. 
These velocities are used to determine the excitation temperature of each $^{13}$CO velocity component, using the accompanying $^{12}$CO spectrum. 
The temperatures are derived under the assumption that the $^{12}$CO emission is in LTE and optically thick at line center.  
Then, for $^{12}$CO, \cite{stahler}

\begin{equation}    
f(T_{kin})=f(T_{ex})=\frac{T_{B_{0}}}{T_{0}} + f(T_{bg}) \lc
\end{equation}

where the excitation temperature (T$_{ex}$) is taken to be equal to the kinetic temperature (T$_{kin}$), T$_{B_{0}}$ is the observed brightness temperature at line center, T$_{0}$ is the equivalent temperature of the transition (5.5K for the J=1 - 0 transition of $^{12}$CO), T$_{bg}$ is the blackbody radiation temperature of the background field (2.7K), and

\begin{equation}
f(T)=\frac{1}{exp(T_{0}/T) - 1} \lp
\end{equation}

\item CO fitting (second pass): Linewidth, and optical depth are derived for each $^{13}$CO emission component using the previously obtained temperatures and center velocities. Subsequently the non-thermal linewidths are obtained using:
\begin{equation}
\sigma_{obs}^{2} = \sigma_{nt}^{2} + \sigma_{th}^{2} \lc
\end{equation} 
where $\sigma_{obs}$ is the observed total linewidth, and $\sigma_{nt}$ and $\sigma_{th}$ are respectively the nonthermal and
thermal components of the linewidth.

\item Molecular Column Densities: Column densities for $^{13}$CO for each component along each line of sight are calculated according to the technique described in \cite{stahler}. H$_{2}$ column densities are estimated using a H$_{2}$/ $^{13}$CO abundance ratio of $7.5\times10^5$ for these galactic clouds.

\item Estimating Atomic Parameters: The center velocity, non-thermal linewidth, and temperature for the HI gas are determined for each velocity component along each line of sight by convolving the previously obtained $^{13}$CO parameters to the GBT 9' beam as weighted by the $^{13}$CO column densities according to:

\begin{equation}
X_{j, comp} = \frac{\sum_{i}{f(\theta_{i,j})\ X_{i,comp}\  N^{^{13}CO}_{i,comp}}}{\sum_{i}{f(\theta_{i,j})\ N^{^{13}CO}_{i,comp}}}  \lc
\end{equation}

where $X_{j, comp}$ represents the temperature, non-thermal linewidth, or center velocity of a particular velocity component at sky position j, i represents all nearby sky positions to be convolved, $f(\theta_{i,j})$ represents the beam response function for angular distance $\theta$ between points i and j, and $ N^{^{13}CO}_{i,comp}$ represents the corresponding $^{13}$CO column density for a particular component at position i. For comparison with HI column densities, the H$_{2}$ column densities are similarly convolved without the mass weighting factor.

\item Having estimates of the temperature, central velocity, and linewidth for each HINSA component we can, for any value of the
  optical depth, remove the HINSA absorption to obtain the background emission using equation \ref{maineqn}. 
Since their true ordering along the line of sight is not known, as a matter of definition velocity components with the highest positive velocities are taken to be the most distant.

\item Finally we search for those values of $\tau_{0}$ and $\sigma_{H}$ which minimize the value of equation \ref{tominimize} 
to arrive at our solutions.

\end{enumerate}
\end{appendix}
\end{document}